\documentclass{emulateapj}\usepackage{apjfonts}\LongTables

\def\wprime{\ensuremath{W^{\prime}}}

\newcommand\z{\ensuremath{z_{850}}}
\newcommand\g{\ensuremath{g_{475}}}

\newcommand\kmsMpc{km~s$^{-1}\,$Mpc$^{-1}$}
\newcommand\etal{{et~al.}} 
\newcommand\mM{\ifmmode(m{-}M)\else$(m{-}M)$\fi}
\newcommand\mMfv{\ensuremath{\Delta(m{-}M)_{FV}}}

\newcommand\hst{{\it HST}}
\newcommand\zacs{\ifmmode z_{850}\else$z_{850}$\fi}
\newcommand\iacs{\ifmmode i_{775}\else$i_{775}$\fi}
\newcommand\gacs{\ifmmode g_{475}\else$g_{475}$\fi}
\newcommand\racs{\ifmmode r_{625}\else$r_{625}$\fi}
\newcommand\vacs{\ifmmode V_{606}\else$V_{606}$\fi}
\newcommand\gz{{\ifmmode{(g_{475}{-}z_{850})}\else$(g_{475}{-}z_{850})$\fi}}
\newcommand\gzacs{{\ifmmode{g_{475}{-}z_{850}}\else$g_{475}{-}z_{850}$\fi}}
\newcommand\riacs{{\ifmmode{r_{625}{-}i_{775}}\else$r_{625}{-}i_{775}$\fi}}
\newcommand\rzacs{{\ifmmode{r_{625}{-}z_{850}}\else$r_{625}{-}z_{850}$\fi}}
\newcommand\izacs{{\ifmmode{i_{775}{-}z_{850}}\else$i_{775}{-}z_{850}$\fi}}
\newcommand\vzacs{{\ifmmode{V_{606}{-}z_{850}}\else$V_{606}{-}z_{850}$\fi}}
\newcommand\vi{{\ifmmode{(V{-}I)}\else$(V{-}I)$\fi}}
\newcommand\Nbar{\ensuremath{\overline{N}}}
\newcommand\Lbar{\ensuremath{\overline{L}}}
\newcommand\Ltot{\ensuremath{L_{\rm tot}}}

\newcommand\Nzbar{\ensuremath{\overline{N}_{z}}}
\newcommand\Nbarz{\Nzbar}
\newcommand\NIbar{\ensuremath{\overline{N}_{I}}}
\newcommand\zbar{\ensuremath{\overline{z}_{850}}}
\newcommand\ztot{\ensuremath{z_{850, {\rm tot}}}}
\newcommand\Ibar{\ensuremath{\overline{I}}}
\newcommand\Hbar{\ensuremath{\overline{H}_{160}}}
\newcommand\zbHb{\ensuremath{\overline{z}_{850}{\,-\,}\overline{H}_{160}}}
\newcommand\mbari{\ifmmode\overline{m}_I\else$\overline{m}_I$\fi}
\newcommand\mbarz{\ifmmode\overline{m}_z\else$\overline{m}_z$\fi}
\newcommand\mbar{\ifmmode\overline{m}\else$\overline{m}$\fi}
\newcommand\mtot{\ensuremath{m_{\rm tot}}}
\newcommand\Mbar{\ifmmode\overline{M}\else$\overline{M}$\fi}
\newcommand\lbar{\ifmmode\overline{L}\else$\overline{L}$\fi}
\newcommand\Mbarz{\ifmmode\overline{M_z}\else$\overline{M}_z$\fi}
\newcommand\lta{\lesssim}
\newcommand\gta{\gtrsim}
\newcommand\cote{C{\^ o}t{\' e}}
\newcommand\jordan{Jord{\'a}n}
\def\vcsi{ACSVCS-I}
\def\vcsii{ACSVCS-II}
\def\vcsiv{ACSVCS-IV}
\def\vcsv{ACSVCS-V}
\def\vcsvi{ACSVCS-VI}
\def\vcsxiii{ACSVCS-XIII}
\def\fcsi{ACSFCS-I}
\def\sigcos{\ensuremath{\sigma_{\rm cos}}}
\def\sigerr{\ensuremath{\sigma_{\rm err}}}

\def\sigest{\ensuremath{\sigma_{\rm est}}}
\def\sigestVir{\ensuremath{{\sigest}_{, \rm Vir}}}
\def\sigestFor{\ensuremath{{\sigest}_{, \rm For}}}

\shortauthors{{Blakeslee et al.}}
\shorttitle{ACS Fornax Cluster Survey. V.}

\begin{document}

\title{The ACS Fornax Cluster Survey. V. 
Measurement and Recalibration of Surface Brightness Fluctuations and 
a Precise Value of the Fornax--Virgo Relative Distance\altaffilmark{1}}

\author{John P. Blakeslee\altaffilmark{2,3},
Andr\'es Jord\'an\altaffilmark{4,5}, 
Simona Mei\altaffilmark{6,7},
Patrick C\^ot\'e\altaffilmark{2}, 
Laura Ferrarese\altaffilmark{2},
Leopoldo~Infante\altaffilmark{5},
Eric~W.~Peng\altaffilmark{8},
John~L.~Tonry\altaffilmark{9},
and 
Michael~J.~West\altaffilmark{10}
}
\altaffiltext{1}{Based on observations with the NASA/ESA 
{\it Hubble Space Telescope} obtained at the Space Telescope Science Institute,
which is operated by the Association of Universities for Research in Astronomy,
Inc., under NASA contract NAS 5-26555}
\altaffiltext{2}{Dominion Astrophysical Observatory,
Herzberg Institute of Astrophysics, National Research Council of Canada, Victoria, 
BC V9E\,2E7, Canada; John.Blakeslee@nrc.ca}
\altaffiltext{3}{Department of Physics and Astronomy,
Washington State University, Pullman, WA 99163-2814}
\altaffiltext{4}{Harvard-Smithsonian Center for Astrophysics, Cambridge, MA 02138}
\altaffiltext{5}{Departamento de Astronom{\'i}a y Astrof{\'i}sica,
Pontificia Universidad Cat{\'o}lica de Chile, Santiago 22, Chile}
\altaffiltext{6}{University of Paris Denis Diderot,  75205 Paris Cedex 13, France}
\altaffiltext{7}{GEPI, Observatoire de Paris, Section de Meudon, 5 Place
J.\ Janssen, 92195 Meudon Cedex, France}
\altaffiltext{8}{Department of Astronomy, Peking University, Beijing 100871, China}
\altaffiltext{9}{Institute for Astronomy, University of Hawaii, Honolulu, HI 96822}
\altaffiltext{10}{European Southern Observatory, Alonso de Cordova 3107, Vitacura, Santiago, Chile}

\begin{abstract}
We present \gz\ color and \z-band surface brightness fluctuations (SBF)
measurements for 43 early-type galaxies in the Fornax cluster imaged
with the \textit{Hubble Space Telescope} Advanced Camera for Surveys.
These are combined with our earlier measurements for Virgo cluster
galaxies to derive a revised, nonlinear calibration of the \z-band SBF
absolute magnitude \Mbarz\ as a function of \gz\ color,
valid for the AB color range $0.8<\gz< 1.6$.  In all, we
tabulate recalibrated SBF distances for 134 galaxies in Virgo, Fornax, 
the Virgo \wprime\ group, and NGC~4697 in the Virgo Southern Extension. The calibration
procedure yields a highly precise relative distance modulus for Fornax
with respect to Virgo of $\Delta(m-M)_{FV} = 0.42\pm0.03$ mag, or a
distance ratio $d_F/d_V = 1.214\pm0.017$.  The resulting Fornax distance
modulus is $(m-M)_{\rm For} = 31.51\pm0.03\pm0.15$ mag, corresponding to
$d_F=20.0\pm0.3\pm1.4$ Mpc, where the second set of error bars reflects
the total systematic uncertainty from our assumed Virgo distance of 16.5~Mpc.
The rms distance scatter for the early-type Fornax cluster galaxies is
$\sigma_d = 0.49^{+0.11}_{-0.15}$~Mpc, or a total line-of-sight depth of 
$2.0^{+0.4}_{-0.6}$~Mpc, consistent with its compact appearance on the sky.
This translates to a depth scatter smaller than the intrinsic, or ``cosmic,''
scatter \sigcos\ in the SBF calibration, unlike the case for the larger Virgo
cluster.  As a result, we are able to place the first tight constraints on the
value of \sigcos.  We find $\sigcos = 0.06\pm0.01$ mag, with a firm upper limit
of $\sigcos<0.08$~mag, for the subsample of galaxies with $\gz>1.02$, but it is
about twice as large for bluer galaxies.  We also present an alternative SBF
calibration based on the `fluctuation count' parameter $\Nbar = \mbar - \mtot$,
a proxy for galaxy mass.  This gives a consistent relative distance but with
larger intrinsic scatter, and we adopt the result from the calibration 
on \gz\ because of its basis in stellar population properties alone.
Finally, we find no evidence for systematic trends of the galaxy distances
with position or velocity (e.g., no current infall);
the Fornax cluster appears both compact and well virialized.  
\end{abstract}

\keywords{galaxies: clusters: individual (Fornax, Virgo)
--- galaxies: distances and redshifts
--- galaxies: elliptical and lenticular, cD
--- large-scale structure of the universe
}

\section{Introduction}
\label{sec:intro}

It is now more than twenty years since Tonry \& Schneider (1988) first quantified the
surface brightness fluctuations (SBF) method for determining extragalactic distances. 
That paper envisioned SBF potentially being used to measure distances to
elliptical galaxies out to 20 Mpc from the ground and being calibrated from
stellar evolution models tied to Galactic clusters with distances from main
sequence fitting.  Tonry \& Scheider noted that the absolute fluctuation
magnitude \Mbar\ would vary with the age and metallicity of a stellar
population, but they predicted that the mean color of a galaxy could be used to
constrain \Mbar\ to within 0.4~mag, allowing galaxy distance measurements
accurate to at least $\sim\,$20\%.

The first major application of the fully developed SBF method was by
Tonry \etal\ (1990), who applied it to a sample of
Virgo cluster galaxies in the $VRI$ bandpasses. They calibrated the \Mbar\ zero
point from measurements in M31 and M32, assuming the Cepheid distance to M31.
The depth of Virgo, including the unexpected discovery that the giant elliptical
NGC\,4365 apparently lay in the background, precluded a good determination of the
dependence of \Mbar\ on stellar population.  Soon afterward, observations of
galaxies in the compact Fornax cluster provided the first fully empirical SBF
calibration, determining $\Mbar_I$ as a function of galaxy \vi\ color (Tonry~1991). 
The ensuing ground-based SBF Survey of Galaxy Distances (Tonry \etal\ 1997,
2001) determined distances to some 300 early-type galaxies and spiral 
bulges.  It also uncovered and corrected various problems with the photometric
consistency of the earlier observations.  The median error in distance modulus
from the ground-based SBF survey was 0.22 mag, or 10\%, including an estimated
``cosmic scatter'' of 0.06~mag intrinsic to the \Mbar--\vi\ relation (Tonry
\etal\ 2000).  This level of accuracy was a substantial improvement over
previous large surveys using other distance indicators.  

At the same time, SBF measurements with the repaired \textit{Hubble Space
Telescope} (\hst) gave the first hints of the enormous potential of the method
with space-based resolution (Ajhar \etal\ 1997, 2001; Pahre \etal\
1999; Neilsen \& Tsvetanov 2000).  However, the characteristics of WFPC2 (small area
of the PC chip; severe undersampling of the WF chips; modest quantum
efficiency) limited the use of this instrument for SBF observations.  For
further details on the first decade of SBF studies, see the review by Blakeslee
\etal\ (1999).  For information on applications of the method using NICMOS
on \hst, including stellar population effects and the unique systematics of
that detector, see Jensen \etal\ (2001, 2003).

More recently, there has been renewed interest in ground-based SBF studies,
mainly targeting dwarf galaxies in several nearby groups using large-aperture
telescopes (e.g., Mieske \etal\ 2003, 2006; Jerjen 2003; Jerjen \etal\ 2004;
Dunn \& Jerjen 2006).  There has also been significant theoretical effort to
predict the behavior of SBF magnitudes in various bandpasses as a function
of stellar population (Liu \etal\ 2000; Blakeslee \etal\ 2001b; 
Mei \etal\ 2001; Cantiello
\etal\ 2003; Mouhcine \etal\ 2005; Raimondo \etal\ 2005; Marin-Franch \&
Aparicio 2006; Lee \etal\ 2007; Cervi\~no et al.\ 2008).   Additionally,
SBF measurements in Magellanic Cloud star clusters of varying ages 
(Gonz{\'a}lez et al.\ 2004; Gonz{\'a}lez-L{\'o}pezlira et al.\ 2005;
Raimondo et al.\ 2005) have provided important new tests and 
calibration data for stellar population synthesis modeling.
These studies show that some discrepancies remain between observations and
models, as well as among the different model predictions, particularly in the
near-IR where thermally-pulsing AGB stars can have a major effect on SBF
magnitudes.  Much work remains to be done to resolve these outstanding issues.

The full promise of the optical SBF method was finally brought to fruition
with the installation of the Advanced Camera for Surveys (ACS) on board \hst.
The ACS Wide Field Channel (WFC) samples the point spread function (PSF) with
a resolution comparable to the WFPC2 planetary camera CCD, but over a much
larger ${\sim\,}3\farcm3{\times}3\farcm3$
field and with about five times the throughput at the
wavelengths typically used for SBF analyses.
SBF investigations with ACS/WFC in the F814W bandpass (most similar to the $I$
band) include the first studies of large samples of high signal-to-noise
radial SBF gradients in early-type galaxies (Cantiello \etal\ 2005; 2007a),
the first optical SBF distance measurements out to $\sim\,$100 Mpc or
beyond (Biscardi \etal\ 2008), and
a precise distance to a peculiar gas-rich S0 galaxy in the Dorado group
(Barber Degraaff \etal\ 2007).  The ACS/WFC has also afforded the first
samples of reliable $B$-band
(F435W) SBF measurements beyond the Local Group (Cantiello \etal\ 2007b),
which are useful for stellar population work.

However, the potential of the SBF method when combined with a large-format,
high-throughput, well-sampled, space-based imager has been most spectacularly
demonstrated by its application in the F850LP bandpass (hereafter \zacs) as
part of the ACS Virgo Cluster Survey (ACSVCS; \cote\ \etal\ 2004, hereafter
\vcsi), a two-band imaging survey of 100 early-type galaxies in the Virgo
cluster with the ACS/WFC.  The data analysis and calibration of the \zacs-band
SBF method are described in detail by Mei \etal\ (2005a,b; hereafter
\vcsiv, \vcsv).  The precision of the ACSVCS SBF distances is about
three times better than for the same galaxies as measured in the ground-based
$I$-band survey, and it has been possible to measure distances for about three
times as many Virgo galaxies.  As a result, the ACSVCS SBF measurements provide the
first clear resolution of the 3-D distribution of the early-type galaxy
population in Virgo (Mei \etal\ 2007; hereafter \vcsxiii).

The ACS Fornax Cluster Survey (ACSFCS; \jordan\ \etal\ 2007, hereafter ACSFCS-I)
was designed as a companion survey to the ACSVCS, with similar goals and
observing strategy, but targeting 43 galaxies in the Fornax cluster, the next
nearest cluster after Virgo.   Other papers in this series deal with the
central brightness profiles in Fornax and Virgo early-type
galaxies and with galaxy scaling relations 
(\cote\ \etal\ 2007, and in preparation).
Future works will address isophotal analysis of galaxy structure
(Ferrarese \etal, in preparation), 
as well as the properties of the globular cluster populations.
As discussed in \fcsi, Fornax is considerably more
compact and regular in shape than Virgo, with a central density of galaxies
about twice as large but a total mass nearly an order of magnitude lower.  The
two clusters therefore make a useful comparison for investigations of
environmental effects on galaxies, a prime motivation for the ACSFCS.  The
compact structure of Fornax has also made it a frequent target for distance scale
studies (e.g., Tonry 1991; Madore \etal\ 1998; Dunn \& Jerjen 2006), but
different methods often disagree regarding the relative distance between the Fornax and
Virgo clusters (e.g., Ferrarese \etal\ 2000).

An important goal of the ACSFCS was therefore to refine the \zacs\ SBF
calibration and take advantage of the homogeneity of the two data sets
and small internal scatter of the method to determine a precise relative
distance between Virgo and Fornax.  Knowledge of the relative distance is
essential for making accurate comparisons between the galaxy and star cluster
properties in these two archetypal clusters.  Here, we present the SBF analysis
of the 43 galaxies observed in the ACSFCS.  The following section summarizes
the observations and data analysis.  \S\,\ref{sec:msr} details our SBF and
galaxy \gz\ color measurements and infers the dependence of F850LP SBF
on color from the Fornax data alone.
\S\,\ref{sec:cal} combines these measurements with our earlier ACSVCS SBF
measurements to determine a calibration from the full sample of galaxies and
a precise estimate of the relative distance modulus between the two
clusters.   In \S\,\ref{sec:disc}, we compare our new measurements 
to literature values, examine the structural properties of Fornax, and 
discuss an alternative SBF calibration based on the parameter \Nbar,
the difference between the SBF magnitude and the total magnitude of the galaxy.
The final section provides a summary of our results.

\section{Observations and Reductions}
\label{sec:obs}

A complete sample of 44 Fornax cluster galaxies was initially targeted as
part of the Cycle~13 \textit{HST} GO program 10217, the ACS Fornax Cluster
Survey.  Of these, 42 constitute a complete sample of early-type galaxies 
with total $B$-band magnitudes $B_T\le15.5$ from the Fornax Cluster Catalogue
(FCC) of Ferguson (1989).  The two other targets, NGC\,1340 and IC\,2006, are
bright early-type galaxies just outside the FCC survey area.  All of the
targets are located within 3\fdg25 of the cD galaxy NGC\,1399, or about
1.1~Mpc for $d\approx20$~Mpc.  However, because of a guide star acquisition
failure, the bright elliptical NGC\,1379 (FCC\,161) was not observed, leaving
a total sample of 43 galaxies.

Full details on the observational program, including motivations, design
specifications, sample properties, observing log, and basic image processing
methods, are given in \fcsi. Further information on the data reduction
techniques is given by \jordan\ \etal\ (2004, hereafter \vcsii).
Briefly, each galaxy was observed for one
orbit, including two exposures totaling 760~s in the F475W (\g) bandpass and
three exposures totaling 1220~s in F850LP (\z).  The exposures were dithered;
we used the alignment routines in the Apsis package (Blakeslee \etal\ 2003) to
determine offsets, then combined the images using the Multidrizzle (Koekemoer
\etal\ 2002) interface to the Drizzle (Fruchter \& Hook 2002) image resampling
software.  We use the Lanczos3 interpolation kernel to reduce the small-scale
noise correlations.  As discussed in detail by \vcsiv, this is an important
consideration for obtaining accurate SBF power spectrum measurements.

The data were photometrically calibrated using the F475W and F850LP zero
points from Sirianni \etal\ (2005).  Although small refinements to these zero
points are available from STScI on the ACS data analysis
webpage\footnote{http://www.stsci.edu/hst/acs/analysis/zeropoints}, we use the
published values to retain consistency with our earlier ACSVCS and ACSFCS
publications.  The revisions to the absolute sensitivity of the ACS detectors
have no bearing on the internally-calibrated SBF results of the present work.
All of the observations were completed before the change of the ACS/WFC
operating temperature in July 2007 and consequent change in
photometric sensitivity.  The photometry was corrected for Galactic extinction
using the dust maps of Schlegel \etal\ (1998) with the extinction ratios 
adopted by \vcsii\ from Sirianni \etal\ (2005).

\vcsiv\ outlined the general steps for measuring the SBF amplitude from the
power spectrum of a galaxy image and used realistic image simulations to
demonstrate that our SBF analysis of the ACS/WFC data produced accurate
results.  The analysis is performed on the F850LP image to measure the SBF
magnitude \zbar\ because the SBF is bright and well-behaved in this bandpass
(\vcsi; \vcsv).  It is too faint to measure reliably in our F475W images;
however, we use both images to measure the galaxy \gz\ color and calibrate
the variation of \zbar\ with stellar population, as described in detail below.
The basic SBF method is the same as described in many previous works
(e.g., Tonry \etal\ 1990; Jacoby \etal\ 1992; Jensen \etal\ 1998; Blakeslee
\etal\ 1999; Mei \etal\ 2003).
It involves constructing a two-dimensional model of the galaxy light,
subtracting the model and sky from the image, detecting sources (stars, globular
clusters, background galaxies), masking the sources and any dust or other
irregular features in the galaxy-subtracted ``residual image,'' measuring the amplitude of
the power spectra in different regions of the masked residual image, correcting the
measurements for contamination from undetected sources, and combining the
results to obtain the average SBF magnitude.

Specific details on the analysis regions, masking, corrections, etc.\ for the
ACSVCS data are given by \vcsv. The method for identifying regions affected by
dust is described in detail by Ferrarese \etal\ (2006, hereafter \vcsvi).
Details on the two-dimensional galaxy fitting for the present sample are
provided in \fcsi.  In general, we followed the same SBF reduction procedures
as in the ACSVCS in order to ensure homogeneity between the two surveys.
Where the analysis was modified, mainly for the sake of streamlining, we
verified that the changes caused no systematic differences in the results.
The following section describes our SBF magnitudes and color measurements,
noting any areas where the procedures have been revised from our earlier work.

\section{SBF and Color Measurements}
\label{sec:msr}

We measure the power spectrum of the masked residual image in a series of
contiguous, concentric annuli with inner radii of 1\farcs6, 3\farcs2,
6\farcs4, 12\farcs8, 19\farcs2, 25\farcs6, 32\arcsec, and 38\farcs4, the same
annuli as used for the ACSVCS measurements. The galaxy center is
determined from our isophotal modeling.  In many cases, the innermost one
or two annuli are omitted because they suffer from poor model residuals or
significant dust contamination.  In addition, we only use annuli where the
mean galaxy surface brightness is at least 60\% of the sky level, which
effectively sets the outermost radius for most galaxies (except the giants,
for which the galaxy surface brightness is significantly above the sky level
for all annuli).  This limit is slightly more conservative than the 50\% value
quoted in \vcsv, but the small change has no significant systematic effect on
the results. The median number of annuli analyzed per galaxy is five (the
average is~5.2).

The power spectrum of each annulus in each galaxy is azimuthally
averaged and modeled as a linear combination of two components:
\begin{equation}
P(k) = P_0 \times E(k) + P_1 \,,
\label{eq:powspec}
\end{equation}
where $P_0$ and $P_1$ are constants and $E(k)$ is the ``expectation power
spectrum,'' which is the convolution of the power spectrum of the normalized
PSF with the mask function of the annular region.  As noted in the previous
section, \vcsiv\ demonstrated that the power spectra are not significantly
modified by pixel correlation induced by our interpolation kernel.  The
coefficient $P_0$ represents the signal we are trying to measure, while $P_1$
includes shot noise, read noise, and any other sources of variance that are
not convolved with the PSF.  

We used two different robust fitting routines available from the IDL Astronomy
Library\footnote{http://idlastro.gsfc.nasa.gov/}
% at Goddard Space Flight Center
to determine the coefficients in equation~(\ref{eq:powspec}).  These were the
least absolute deviation procedure LADFIT used in \vcsv, and an iterative
rejection procedure using bisquare weights called ROBUST\_POLY\_FIT.  The
median difference in the final results using these two different fitting
methods was 0.006~mag, with an rms scatter in the differences of 0.034~mag.
Because there was no significant systematic difference, we averaged the results
from the two sets of fits and included half the difference in quadrature in
our error estimation.
% Notes on robust_poly_fit:
%
% For the initial estimate, the data is sorted by X and broken into 
% NDEGREE+2 sets. The X,Y medians of each set are fitted to a polynomial
% via POLY_FIT.   Bisquare ("Tukey's Biweight") weights are then 
% calculated, using a limit  of 6 outlier-resistant standard deviations.
% The fit is repeated iteratively until the robust standard deviation of 
% the residuals changes by less than .03xSQRT(.5/(N-1)).

The final step in determining the SBF signal for each annulus is to estimate
the residual variance contamination $P_r$ due to undetected sources and
subtract it from $P_0$ to derive the variance due to SBF: $P_{\rm SBF} = P_0 -
P_r$.  This is normalized by the galaxy surface brightness and converted to
the SBF magnitude~\zbar.  We used the same software and followed the identical
procedures as in \vcsv\ to estimate the background variances based on the
expressions given by Blakeslee \& Tonry (1995).  Because the data go well
beyond the turnover in the globular cluster luminosity function, the median
correction for the full sample of galaxies was only 0.024~mag, with a range
from 0.006 to 0.08~mag.  As in \vcsv, we assign a 25\% uncertainty to this
correction.

\begin{figure}\epsscale{1.1}
\plotone{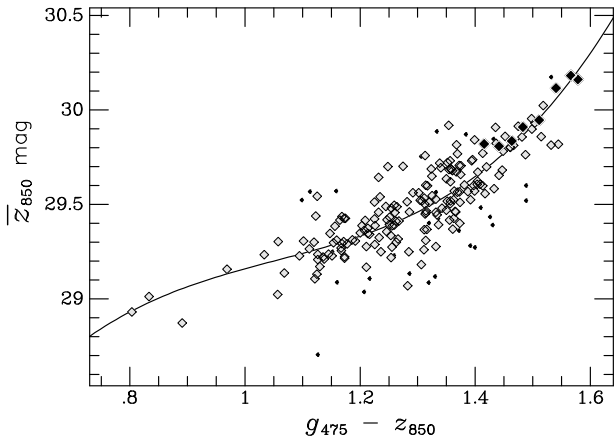}
\caption{SBF \zbar\ magnitude as a function of \gz\ color for
  individual annuli (diamonds) within the ACS Fornax Cluster Survey galaxies.
  The curve is a third-order polynomial fit to the full set of annuli after
  median-smoothing; it is used to reject annuli that are 3-$\sigma$
  outliers with respect to the shifted relation defined by each galaxy's
  set of annuli.  The small points show the rejected annuli.  The large
  solid diamonds highlight the annuli of NGC\,1399 (FCC213), the galaxy
  with the largest number of usable annuli.
\label{fig:sbfannuli}}
\end{figure}

For each galaxy, we then have a set of corrected \zbar\ measurements over
different radial ranges.  We measure the galaxy colors in the same
annuli, with the same masks applied, and estimate the errors based on
detector and shot noise, sky uncertainties in each bandpass (\fcsi), and
an assumed 1\% flat-fielding positional error in the photometry (e.g.,
Sirianni \etal\ 2005).  Figure~\ref{fig:sbfannuli} shows
the measurements for all annuli.  As in \vcsv, we find that \zbar\ becomes
fainter as \gz\ gets redder, and the relation steepens at the red end.  We can
use this relation to reject discordant annuli before averaging.  We did this by
first median filtering the relation using steps of 0.04~mag in \gz, then
fitting a cubic (third-order) polynomial to the median filtered relation.  The curve in
Figure~\ref{fig:sbfannuli} shows the result.

We assume that the annuli in each galaxy should follow the same general
relation, but with a shift depending on the galaxy distance, and we used this
to reject any annuli more than 3$\sigma$ discrepant from the shifted relation
for each galaxy.  This process eliminates 15\% of the galaxy annuli, or an
average of about one annulus per galaxy.
% modified:
The rejected annuli, shown as small dots in Figure 1, are usually the
innermost or outermost ones.  They are non-Gaussian outliers that may be
affected by poor model residuals or imperfect masking of chip defects,
real objects, or clumpy dust.
A similar culling of
annuli was conducted for the ACSVCS SBF analysis, and we have simply made it
more automated.  Following this, we average the measurements for the remaining
annuli to obtain a single value of \zbar\ and of \gz\ for each galaxy.

We could instead use the calibration effectively defined by the annuli to obtain a
relative distance modulus for each annulus, then average these to obtain a
single distance modulus for each galaxy.  We have tested this approach, and the
final results are insensitive to whether or not the averaging is done on the
magnitudes and colors or on the distance moduli, as long as the same
calibration relation is used.  This is because the calibration is quite linear
over the color range of any individual galaxy.  For instance, we have
highlighted in Figure~\ref{fig:sbfannuli} the annuli for NGC\,1399, the galaxy
with the most annuli remaining after the culling process.  We chose
to average SBF magnitudes and colors before applying the calibration 
in order to follow the same
procedure as in \vcsv, and to simplify the derivation of a single, combined
Virgo+Fornax SBF calibration, a primary goal of this investigation.

%%%%%%%%%%%%%%%%%%%%%%%%%%%%%%%%%%%%%%%%%%%%%%%%%%%%%%%%%%%%%%

\begin{figure}\epsscale{1.1}
\plotone{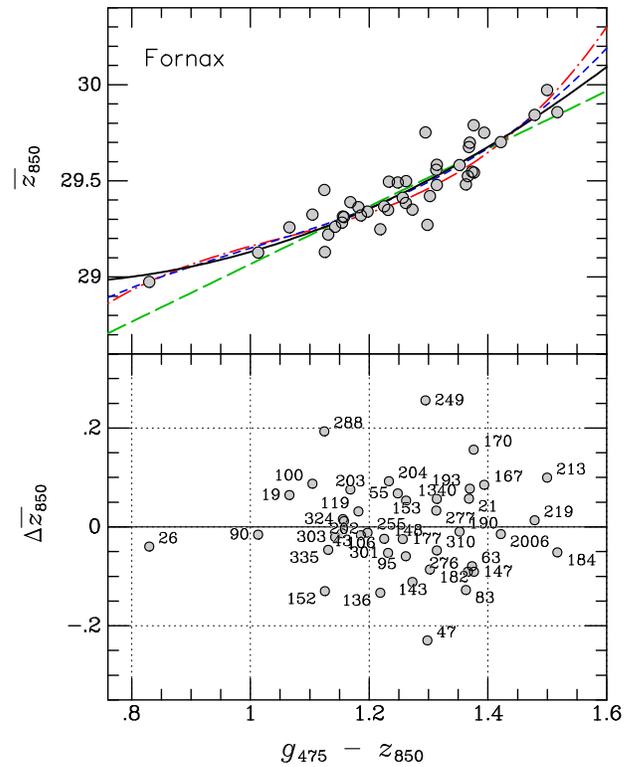}
\caption{Mean SBF \zbar\ magnitude vs \gz\ color for the 43 ACS Fornax Cluster
Survey galaxies.  The top panel shows the linear (green dashed line),
quadratic (solid black curve),
and cubic (blue dashed curve)
polynomial fits to these averaged data, 
as well as the cubic polynomial fit to the
annuli from Figure~\ref{fig:sbfannuli} (red dot-dashed curve).
The quadratic and cubic fits to these data
have very similar RMS scatters of 0.092 and 0.091 mag, respectively. 
The lower panel shows the residuals with respect to the quadratic fit;
the galaxies are labeled by their FCC numbers.
\label{fig:sbffornax}}
\end{figure}

Figure~\ref{fig:sbffornax} (top panel) plots the averaged \zbar\ and \gz\ values for our
43 ACSFCS galaxies.  The numbers appear in Table~\ref{tab:fornax}, along with
the final distances derived in \S\,\ref{sec:cal} and the common names for
the galaxies.  Magnitudes and colors are all corrected for Galactic
extinction and reddening as described in \S\,\ref{sec:obs}.
These data
allow us to define a calibration for the variation in apparent SBF \z-band
magnitude on galaxy color based entirely on Fornax galaxies, analogous to that
derived in \vcsv\ for Virgo.
Fitting a linear relation to the full sample of 43 Fornax galaxies, we find
\begin{equation}
\zbar \;=\; (29.517\pm0.016) \,+ \,(1.50\pm0.15)\,[\gz{-}1.3] \,,
\label{eq:lincal}
\end{equation}
where the errors are determined from bootstrap resampling (e.g., Press \etal\
1992), and the rms scatter in the fit is 0.10~mag.  This relation is plotted
in Figure~\ref{fig:sbffornax}.  It is interesting to compare
% this result
equation~(\ref{eq:lincal})
with the corresponding linear fit for Virgo from equation~(14) of
ACSVCS-V. The slopes are consistent within the combined errors, and the zero
points differ by $0.43\pm0.03$~mag.  This is a measure of the relative
distance modulus between Fornax and Virgo, to the extent that a linear fit
provides an accurate description of the \zbar--\gz\ relation.  However,
ACSVCS-V concluded that the linear fit was inadequate.
Similarly, we find that the relation for Fornax exhibits obvious
curvature.  The slope depends on the color range of the galaxies fitted, and
if we omit the single galaxy at $\gz<1$, the slope increases to~1.60.

We also show second- and third-order polynomial fits to the \zbar-\gz\
relation in Figure~\ref{fig:sbffornax} and compare the fitted relation for the
annuli from Figure~\ref{fig:sbfannuli}.  For this sample of SBF measurements,
a quadratic fit provides an adequate description to the curvature in the
relation, and higher order polynomials do not significantly decrease the rms
scatter.  The relation is poorly constrained for $\gz<1$, where there is only a
single galaxy.  The lower panel of Figure~\ref{fig:sbffornax} shows the
residuals with respect to the quadratic relation.  The rms scatter is
0.092~mag, and the largest outlier, at 0.26~mag or (2.8\,$\sigma$), is FCC249.
The next largest outlier is FCC47 at just under 2.5\,$\sigma$.  

The tight relation between the Fornax \zbar\ magnitudes and \gz\ colors
allows us to obtain, for the first time, a reasonably well constrained
 estimate of the intrinsic, or ``cosmic,'' scatter
in this relation.  We can estimate the expected depth of Fornax using the rms
positional scatter of the sample galaxies on the sky.  We find a scatter of
1\fdg462 in right ascension and 1\fdg335 in declination.  We estimate the
rms depth in magnitudes as
\begin{equation}
\sigest \;=\; \sqrt{\frac{1}{2}(\sigma^2_{\rm RA} + \sigma^2_{\rm Dec})} \,
\times \frac{\pi}{180} \times \frac{5}{\ln{10}} \;\hbox{mag}\,,
\label{eq:scat}
\end{equation}
where $\sigma_{\rm RA}$ and $\sigma_{\rm Dec}$ are the rms positional scatters
in degrees of arc.   We find $\sigest = 0.053$~mag for Fornax.  The
mean observational error in distance for our sample due to the combined errors in \zbar\
and \gz\ is 0.047~mag (the median is 0.042~mag).  We can therefore estimate
the cosmic scatter in the \zbar-\gz\ relation as
\begin{eqnarray}
\sigcos &\approx& \sqrt{\sigma_o^2 - \sigest^2 - \sigma_{\rm err}^2} \\
                 &\approx& 0.060 \;\hbox{mag} \nonumber\,,
\end{eqnarray}
where $\sigma_o=0.092$ mag is the observed scatter in the relation, and
$\sigma_{\rm err}$ is the typical measurement error.  Thus, unlike in Virgo,
the estimated depth of the cluster is less than the intrinsic scatter in
the method.  Of course, the value of \sigcos\ is still somewhat
degenerate with assumptions about the cluster depth.
We examine these issues through a more careful
$\chi^2$ analysis in the following section, and discuss 
specific limits on cosmic scatter and cluster depth in \S\,\ref{ssec:struct}.

\section{Combined SBF Calibration and the Fornax-Virgo Relative Distance}
\label{sec:cal}

% Fig 3 - color distribution

\begin{figure}[h]\epsscale{1.05}
\plotone{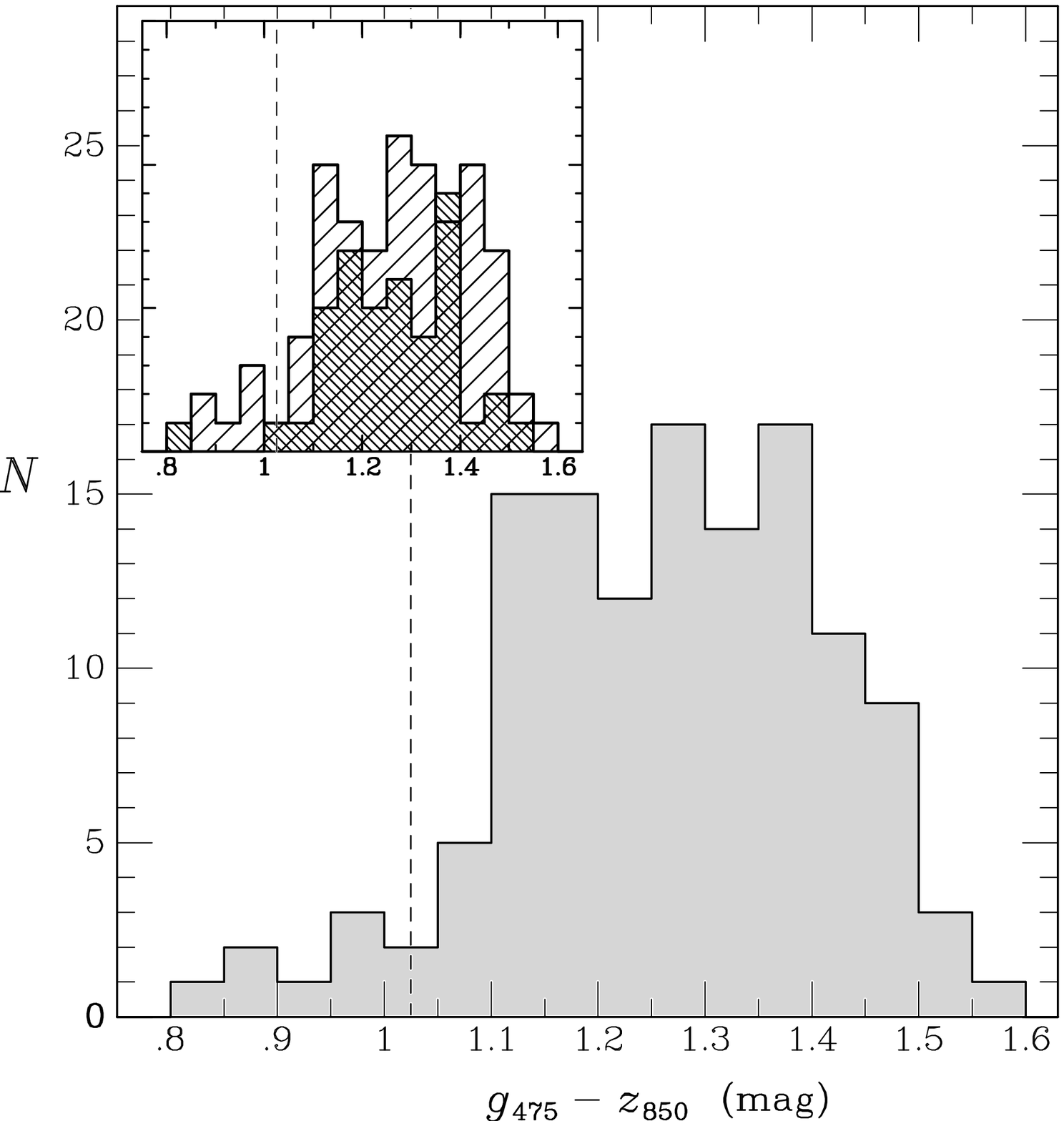}
\caption{Galaxy \gz\ color histogram for the combined sample of 128 
Fornax and Virgo galaxies with SBF measurements.  The dashed line near 
$\gz=1.02$ marks where the scatter in the SBF-color relation increases
for bluer galaxies.  Overall, the mean color is $\gz=1.26\pm0.01$ and the rms 
dispersion is 0.15 mag; for the subsample omitting the blue tail, the 
mean and dispersion are $1.28\pm0.01$ and 0.12 mag, respectively.  The inset
shows the Virgo (broad hatching) and Fornax (narrow hatching) histograms
plotted separately.  The vertical ticks in the inset mark the same
intervals as for the larger plot.
\label{fig:gzhisto}}
\end{figure}

We now combine our ACS Fornax and Virgo survey measurements to obtain an
improved calibration.  We first look at the color distributions of the galaxies
in the two samples.  Figure~\ref{fig:gzhisto} displays the \gz\ histogram for
the total sample of 128 galaxies in Fornax and Virgo (omitting the more
distant \wprime\ galaxies) with SBF measurements from the present work and
from \vcsxiii, respectively.  The overall median color is 1.27~mag and the
mean is $\langle\g{-}\z\rangle = 1.26\pm0.01$, which is the same as for
the Fornax and Virgo samples individually (Figure~\ref{fig:gzhisto} inset).
The color distribution is somewhat broader for Virgo, which has an
rms scatter $\sigma_{g{-}z} = 0.16\pm0.01$ mag, compared to $\sigma_{g{-}z} =
0.13\pm0.02$ mag for Fornax.  This is partly because there are more
massive red ellipticals in Virgo, but also the Virgo sample is
not complete and goes $\sim\,$1~mag deeper in terms of absolute $B$ magnitude
than the Fornax sample; it thus includes lower mass, bluer galaxies.
The color distribution is asymmetric with a tail
to bluer colors.  Omitting galaxies with $\gz<1.02$, the mean and dispersion
are respectively 1.29 and 0.13 for Virgo, and 1.27 and 0.11 for Fornax.  Thus,
the two samples have very similar mean colors, but the dispersion is slightly
larger for Virgo.

\subsection{The Combined Calibration}

Given that the two samples are fairly similar, combining them and fitting for
a single SBF-color relation and relative offset in distance is a reasonable
approach.  To this end, we minimize the value of $\chi^2$ calculated as
\begin{equation}
\chi^2 \;=\;
\sum_i^{N_V} \frac{\left[p_n(x_i) - {\zbar}_{,i}\right]^2}{\,\sigma_{{\rm err,}i}^2 +
  \sigma^2_{\rm cos} + {\sigest^2}_{, \rm Vir}} \,+\, 
\sum_j^{N_F} \frac{\left[p_n(x_j) + \Delta - {\zbar}_{,j}\right]^2}{\sigma_{{\rm err,}j}^2
  +  \sigma^2_{\rm cos} + {\sigest^2}_{, \rm For}}  \,,
\label{eq:chisum}
\end{equation}
where the two sums are over the $N_V$ and $N_F$ galaxies 
in Virgo and Fornax, respectively,
$p_n$ is a polynomial of order $n$, $x_i = {\gz}_i - 1.3$, $\Delta = \mMfv$ is the
relative distance modulus between Fornax and Virgo, $\sigma_{{\rm err,}i}$ is the
total observational error for each galaxy measured perpendicular to the
polynomial (based on error estimates from \S\,\ref{sec:msr}),
\sigcos\ is a fixed value for the cosmic scatter in
the calibration, $\sigestVir$ is the rms depth in mag of Virgo,
and $\sigestFor$ is the rms depth in mag of Fornax.  In practice, we vary
$\Delta$ over a range of values and refit the polynomial calibration at each value.
We omit the five Virgo \wprime\ galaxies from the $\chi^2$ sum, as these are
located 6.5 Mpc beyond the Virgo core (\vcsxiii).  

From equation~(\ref{eq:scat}) above, we estimated the rms depth of Fornax to
be $\sigestFor = 0.053$~mag.  Doing the same calculation for the Virgo
sample gives  $\sigestVir = 0.085$ mag.  Thus, the expected depth
scatter is about 40\% less for Fornax, which is a prime motivation for its
frequent use as a distance method calibrator.  These numbers agree well with
the rms depths of 0.055~mag and 0.082~mag adopted by Tonry \etal\ (2000) for
Fornax and Virgo, respectively, based on smaller samples of galaxies.
[The significantly larger value quoted for the expected rms depth of Virgo in \vcsv\ and
 \vcsxiii\ resulted from the omission of the factor of $\frac{1}{\sqrt{2}}$
in the evaluation of equation~(\ref{eq:scat}).]

In practice, we have found from the $\chi^2$ analysis that the assumption of
a single fixed value for \sigcos\ is a good approximation only at colors
$\gz\gta1.02$ mag, similar to the conclusions of \vcsv.  Minimizing $\chi^2$ for
the subsample of 119 galaxies (77 in Virgo, 42 in Fornax) with $\gz>1.02$, we
find that a second-order polynomial gives the best value of the reduced
$\chi^2$, and that this value reaches $\chi^2_N=1.0$ for the best-fit model
when $\sigcos=0.064$ mag.  This value of \sigcos\ is more dependent on the
Fornax galaxies because \sigcos\ is larger than the magnitude depth of Fornax
but smaller than the depth of Virgo.

There is one galaxy in Fornax (FCC249) that is a non-statistical, 3.5-$\sigma$
outlier from the best-fit combined calibration.  Like the \wprime\ galaxies in
Virgo, it may simply be more distant than the rest of the cluster.  The next
largest outlier in Fornax is 2.6\,$\sigma$.  If we omit FCC249 and refit the
second-order calibration for galaxies with $\gz>1.02$, we obtain
$\chi^2_N=1.0$ for the best-fit model when $\sigcos=0.059$ mag, very similar
to the result for our more approximate calculation in the preceding section
based on Fornax alone.  In \vcsv, a value of $\sigcos=0.05$ was assumed, but
no constraints were possible, given that the assumed depth of Virgo in that
paper was more than twice as large.  Formally, the value of the total $\chi^2$
changes by 1.0 when \sigcos\ is changed in our analysis by only 0.001~mag.
This reflects the precision of the ACS SBF measurements and the small depth of
Fornax.  However, the depth of Fornax is uncertain by at least 15\%, based on
bootstrap resampling of the galaxy positions used in the calculation of
equation~(\ref{eq:scat}) and could be larger if the cluster were elongated
along the line of sight.  If we therefore assume a depth uncertainty of 20\%,
then we estimate values of the cosmic scatter and its 1-$\sigma$ uncertainty
of $\sigcos \approx 0.06\pm0.01$ mag, which we henceforth adopt for galaxies with
$\gz>1.02$~mag.

\begin{figure}\epsscale{1.05}
\plotone{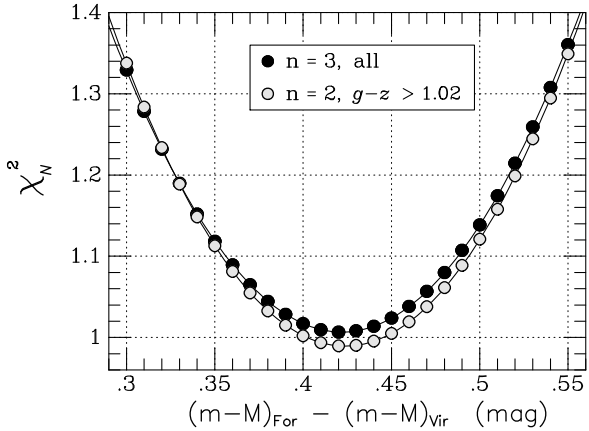}
\caption{%
Reduced $\chi^2$ as a function of the relative Fornax-Virgo distance
modulus \mMfv\ for a 
second-order polynomial fit to the combined \zbar-\gz\ relation for
118 Fornax and Virgo galaxies with \hbox{$\gz>1.02$} (gray circles) and for a 
third-order polynomial fit to 127 Virgo and Fornax galaxies over the full
color range (filled black circles).
The number of free parameters is equal to the order of the polynomial
plus one (the relative distance modulus).  Thus, the $n=2,3$ order
fits have 115 and 123 degrees of freedom, respectively.  In both
cases, we use allowances for cluster depth as described in the text,
and a ``cosmic,'' or internal, scatter in the method of 0.06~mag
(adjusted to give $\chi^2_N\approx1.0$) for the galaxies with
$\gz>1.02$.  The eight bluer galaxies require a larger internal
scatter of $\sim\,$0.13~mag.
The value of $\chi^2$ is computed here at increments
of 0.01~mag, and the curves show that a quadratic provides
excellent fits, with minima at $\mMfv{\,=\,}0.424\pm0.020$ ($n=2$) and
$\mMfv{\,=\,}0.422\pm0.019$ ($n=3$). \hfil
\label{fig:chi}}
\end{figure}

Figure~\ref{fig:chi} shows the values of the reduced $\chi^2$ as a function of
the relative Fornax-Virgo distance modulus $\mMfv$ (represented by $\Delta$ in
equation~\ref{eq:chisum}).  The gray circles show the $\chi^2_N$ values
for the quadratic polynomial, using galaxies with $\gz>1.02$.  The best-fit
relative distance modulus is $\mMfv = 0.424\pm0.020$~mag.  The identical
error~bar is
obtained from the $\chi^2$ analysis as from bootstrap resampling.  The solid
black circles in the figure show  $\chi^2_N$ values
for a cubic polynomial fit to the full range of galaxy colors.  The cubic
polynomial gives a small but significant improvement in this case;
higher-order polynomials are not warranted.  However, in order to obtain
$\chi^2_N\approx1.0$, as is Figure~\ref{fig:chi}, it is necessary to use a
larger intrinsic scatter of $\sigma_{\rm cos, blue}=0.13$ mag for the blue
tail of galaxies at $\gz<1.02$.  This is mostly driven by the scatter for the blue
Virgo galaxies.  The best-fit relative distance modulus in this case is 
$\mMfv = 0.422\pm0.019$, virtually unchanged.   We note that if we include the
galaxy FCC249 and repeat the analysis, the best-fit relative distance modulus
increases by 0.005~mag.

\begin{figure}\epsscale{1.1}
\plotone{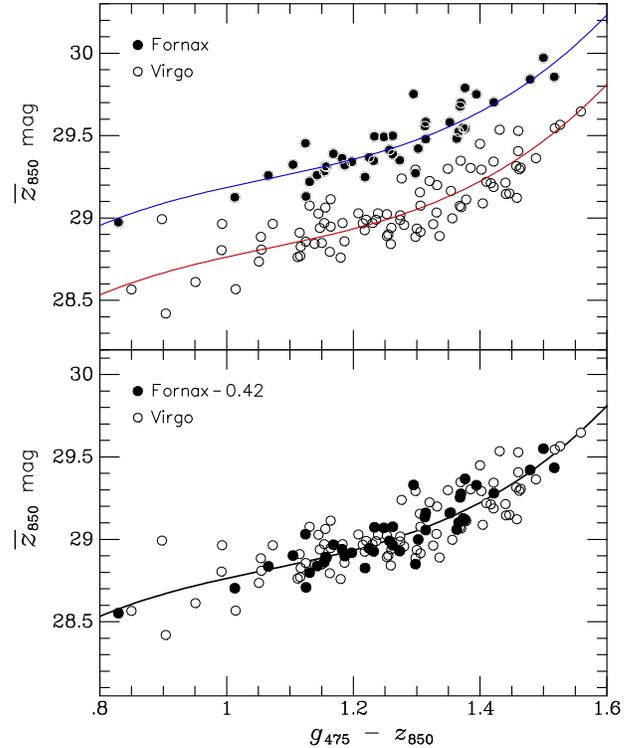}
\caption{SBF \zbar\ magnitude versus \gz\ color for our ACS Virgo and Fornax
  cluster survey galaxies.  The top panel shows the cubic polynomial
  fit to the full sample of galaxies (omitting the \wprime\
  galaxies, not plotted). 
  The relative Fornax-Virgo distance modulus \mMfv\ is 
  a free parameter, and the fit is done by minimizing $\chi^2$,
  including errors in both coordinates.  The
  lower panel shows the two samples shifted together by subtracting
  the best-fit value $\mMfv=0.42\pm0.02$ mag
  from the Fornax galaxy distance moduli. \\
\label{fig:finalcal}}
\end{figure}

The combined cubic polynomial calibration is presented in
Figure~\ref{fig:finalcal}.  The top panel shows the Virgo and Fornax galaxies
prior to shifting; the bottom panel shows the clusters shifted together
by the best-fit $\mMfv$.  Assuming a distance modulus of 31.09~mag for the
Virgo cluster as in \vcsxiii, we determine a final calibration for the
absolute \z\ SBF magnitude of
\begin{equation}
\overline M_z \;=\; -2.04 \,+\, 1.41x \,+\, 2.60x^2 \,+\, 3.72x^3 \,,
\label{eq:cal}
\end{equation}
where $x \,\equiv \,\gz - 1.3$, and equation~(\ref{eq:cal}) is valid for 
$0.8\lta\gz\lta1.6$.  The smooth, continuous variation of \Mbarz\ with
\gz\ indicates there is no dwarf-giant dichotomy with respect to stellar
population properties, apart from an increased scatter for the bluest galaxies.
The precise form of the calibration is uncertain for $\gz\lta1$~mag
because of the larger scatter and low number of points.
The statistical error on the \Mbarz\ zero point in equation~(\ref{eq:cal}) is
0.01 mag, but the systematic uncertainty due to the distance of Virgo is
$\sim\,$0.15 mag, based on the zero-point uncertainty of the Cepheid distance
scale (Freedman \etal\ 2001; Macri \etal\ 2006) and the tie between the
Cepheid and early-type galaxy (SBF) distance scales (Tonry \etal\ 2001;
hereafter Ton01). 

The resulting Fornax distance modulus is 
$(m-M)_{\rm For}=31.51\pm0.02\pm0.15$ mag,
corresponding to a distance $d_F=20.0\pm0.2\pm1.4$~Mpc.  
Based on their positions in the sky, the three-dimensional distance between
the center of Virgo and the center of Fornax is then 33.4~Mpc.
Table~\ref{tab:fornax} lists our final galaxy distances calibrated using
equation~(\ref{eq:cal}), and Table~\ref{tab:virgo} presents the data and
homogeneously calibrated distances for the ACSVCS galaxies, plus
NGC\,4697, the dominant elliptical member
of a group in the Virgo Southern Extension (e.g., Tully 1982).
In all, we tabulate new or recalibrated \z-band SBF distances for 134 galaxies.
The NGC\,4697 observational details were identical to those of the
ACSVCS galaxies and have been discussed by \jordan\ \etal\ (2005).
The distance errors in Tables~\ref{tab:fornax} and \ref{tab:virgo}
are the quadrature sums of the
measurement errors and the cosmic scatter $\sigcos$, which is 0.06~mag for
galaxies with $\gz>1.02$, and approximated as twice this for bluer galaxies.

\subsection{Possible Systematics}
% 1. extinction
% 2. cluster-to-cluster zeropoint
% 3. ACS sensitivity

There are a few systematic differences between the Virgo and Fornax samples
that could potentially affect the relative distance estimate.  One is the
mean $E(B{-}V)$ extinction correction: 0.028 mag for Virgo versus 0.013 mag
for Fornax (Schlegel \etal\ 1998).  The $E(B{-}V)$ corrections for the
cD galaxies M87 and NGC\,1399 are 0.022 and 0.013 mag, respectively.
From Schlegel \etal, the error on the extinction estimate is $\pm$16\% of the
value itself.  Because of the dependence of \Mbarz\ on \gz\ color, and
the total-to-selective extinction ratios for \g\ and \z, the error on
\mM\ using equation~(\ref{eq:cal}) scales as $\delta(m-M) \approx -1.7\,
\delta E(B{-}V) \approx \pm0.27\,E(B{-}V)$.  Because the mean extinctions
toward Virgo and Fornax are both low, the quadrature sum of their distance
errors from extinction is only $\pm$0.008~mag.
%, which is negligible when added in quadrature to the 0.02~mag
% error estimate from the $\chi^2$ analysis.

Another difference is in the date of the observations.  The Fornax data were
on average taken 1.5~yr after the Virgo data, and there is some evidence for
small changes in the ACS/WFC photometric sensitivity during this period.
Bohlin (2007) finds a decline in sensitivity of $0.002\pm0.001$ mag~yr$^{-1}$
for F475W and essentially no change for F850LP.  Because of the color term in
the \zbar\ calibration and the 1.5 yr baseline, this translates to an error 
of $-$0.004 mag in \mMfv\ (i.e., the value is underestimated).
However, using a different technique, Riess (2004)
earlier reported evidence for a sensitivity decline in F850LP 
similar to Bohlin's results at shorter wavelengths. If the degradation is
independent of wavelength, then accounting for the effects on both \zbar\ and
color, the error in \mMfv\ would be $\sim {+}0.004$ mag (the value is overestimated). 
Thus, we estimate the systematic uncertainty from this effect to be $\pm0.004$~mag.
All of the data were taken before July 2006 when a significant change in
sensitivity occurred because of a change in operating temperature.

%%% Possible zero-point shift?
One other potential difference is in the absolute zero point of the \zbar\
versus \gz\ relations for Fornax and Virgo.  Is the calibration
truly universal, or could there be small cluster-to-cluster variations?
Based on our analysis and the fit presented in Figure~\ref{fig:finalcal}, the
clusters appear to define a single relation.  However, while there is
an explicit correction for stellar population based on \gz,
it is possible that systemic differences in the population
parameters at a fixed color could cause small shifts in the 
zero point.  This is because age and metallicity variations are not 
\textit{exactly} degenerate in their effects on the relation between \zbar\
and \gz.
For example, it could be that star formation began later in Fornax than in
Virgo, so that the mean galaxy age is younger even at the same color.  
To assess the implications of this possibility, we selected subsets 
of the composite stellar
population models described by Blakeslee \etal\ (2001) and shown in \vcsi.
In one test, simulating a delay of several Gyr in the onset of galaxy
formation, we selected only models where the old stellar component was at
least 3 Gyr younger than the maximum stellar age in the \vcsi\ models.
This selection removed 56\% of the models.  We then refit the
calibration.  The scatter in the model \zbar\ versus \gz\ relation went from
0.066~mag as reported in \vcsi\ to 0.060~mag, our best empirical estimate
for the intrinsic scatter in Fornax.  The shift in zero point was 0.019 mag.

We tried other age selections that approximately preserved the overall color
range but removed up to 79\% of the composite models.  The maximum zero-point
shift from these tests was 0.026~mag, but more typically was $\lta0.02$ mag.
While an exhaustive treatment might involve analyzing the SBF relations in
multiple clusters from detailed semi-analytic simulations (which currently do
not include SBF predictions), we conclude that 0.02~mag is a reasonable
estimate of this systematic zero-point uncertainty.
We add this in quadrature with the 0.02~mag statistical uncertainty and 
the small effects from extinction and ACS photometric sensitivity as
described above.  The total error on \mMfv\ is then $\pm\,$0.03~mag, and
the Fornax distance is $d_F=20.0\pm0.3\pm1.4$~Mpc.  

%%%%%%%%%%%%%%%%%%%%%%%%%%%%%%%%%%%%%%%%%%%%%%%%%%%%%%%%%%%%%%%%%%%%

\subsection{Past Estimates of the Relative Distance}

Before moving on to the discussion in the following section, we note
that our result for the Fornax-Virgo relative distance modulus $\mMfv =
0.42\pm0.03$ mag, which includes all systematic uncertainties,
is well within the range of previous estimates.  For
instance, Ferrarese \etal\ (2000) tabulated values including
$0.09\pm0.27$ mag from the globular cluster luminosity function,
$0.30\pm0.10$ mag from the planetary
nebula luminosity function, and $0.40\pm0.06$ mag from ground-based
$I$-band SBF.  The difference of the weighted averages of the full
samples of tabulated ``good'' SBF distances for Virgo and Fornax in
Ton01 gives $\mMfv= 0.36\pm0.05$ mag.  The fundamental plane analysis by
Kelson \etal\ (2000) implies $\mMfv= 0.52\pm0.17$ mag.  The one
discrepant method among these appears to be the GCLF, suggesting an
intrinsic difference in the mean luminosity of the globular clusters in
these two systems (Blakeslee \& Tonry 1996).  This issue will be
discussed in detail by another paper in our series.

The final \hst\ Key Project Cepheid distances from Freedman
\etal\ (2001) imply  $\mMfv\ = 0.47\pm0.21$ mag.  The relatively large
error bar is mainly due to the spread of 0.54 mag for the three
Cepheid galaxies in Fornax.  There is also good evidence that the
Cepheid galaxies in both clusters avoid the core regions where
early-type galaxies prevail.  \hbox{Despite} their many virtues, Cepheids are
therefore far from ideal for gauging the distances to galaxy clusters.
\hbox{Although} our relative Fornax-Virgo distance is consistent with many
previous studies, its precision is far greater.  Because of the 
homogeneity of the ACSVCS and ACSFCS observations and SBF analyses,
we believe that this level of precision reliably represents the accuracy of
our relative distance measurement.

\section{Discussion}
\label{sec:disc}

Our combined sample of SBF measurements for 134 early-type galaxies in the 
Virgo and Fornax clusters, Virgo \wprime\ group,
and NGC\,4697 in the Virgo Southern Extension
constitutes the largest homogeneous set of SBF distances available.  All
were observed with the ACS/WFC for one orbit in the F850LP and F475W bandpasses.
The ground-based $I$-band survey
(Tonry \etal\ 1997; Ton01) of 300 galaxies was an ambitious, decade-long
project, but heterogeneous in terms of seeing conditions, detectors, $I$-band
filter characteristics, and telescopes.  We now compare our results to those
of some previous Fornax and Virgo SBF studies before going on to examine the
questions of the structure of Fornax and a possible alternative
calibration for our \z-band SBF data.

\subsection{Other Optical SBF Studies}
\label{ssec:lit}

\begin{figure}\epsscale{1.0}
\plotone{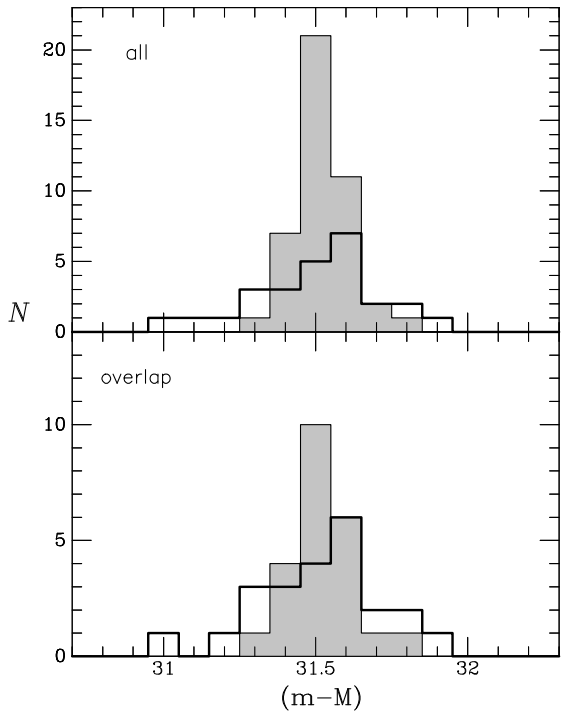}
\caption{Fornax SBF distance moduli from our ACSFCS sample
  (gray histograms) and from the ground-based survey of Tonry et al.\
  (2001; thick-lined open histograms).  Top panel shows the histograms
  for the full samples of 43 ACSFCS and 26 ground-based measurements; 
  lower panel shows histograms for the subsample of 23 galaxies in
  common between the two surveys (note different vertical ranges).
\label{fig:mMhistos}}
\end{figure}

There have been a number of studies of the SBF properties of Fornax cluster
galaxies in both optical and near-infrared passbands.  The largest compilation
of Fornax SBF data prior to our ACS survey was the subset of 26 ground-based
$I$-band SBF measurements in Fornax from Ton01.  The ACSFCS includes 23 of
these galaxies.  (Aside from NGC\,1379, for which the \hst\ observation
failed, Ton01 observed NGC1366, which is outside the FCC survey area, and
NGC1386, which is classified as Sa in the FCC.)  Figure~\ref{fig:mMhistos}
compares the histograms of Fornax SBF distances from the two surveys.  The top
panel includes all Fornax galaxies in each survey; the lower panel includes
the 23 galaxies in common.

As evident in Figure~\ref{fig:mMhistos}, the two surveys agree well in the
mean, but our distance distribution is much tighter.
The median distance modulus for our full sample of 43 galaxies is 31.51 mag,
and the observed rms dispersion is 0.092 mag.  As discussed previously, the
dispersion includes contributions from measurement error, intrinsic scatter,
and Fornax depth.  In comparison, the median of the 26 Ton01 distances is
31.52 mag and the rms dispersion is 0.21~mag.  For the 23 galaxies in
common, the median distance moduli again agree very closely and the rms dispersions
are 0.11 mag and 0.20 mag for ACSFCS and Ton01, respectively.  Allowing for
the spread due to the estimated depth of 0.053 mag, our distances are more accurate by
better than a factor of two, and the measurement errors are smaller by a factor
of~three (but the intrinsic scatter of $\sim\,$0.06 mag is common to both).

\begin{figure}\epsscale{1.05}
\plotone{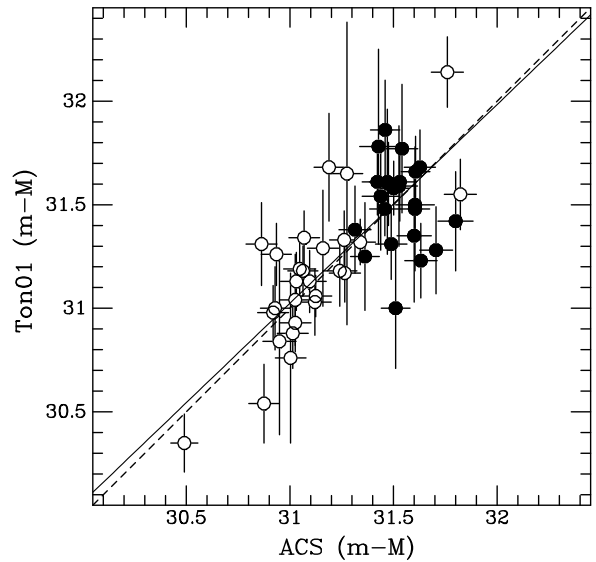}
\caption{Comparison of the ground-based SBF distance moduli from Tonry
  \etal\ (2001) with our SBF distance moduli from the ACS Virgo (open
  circles) and Fornax (filled cirles) cluster surveys.  The solid line
  shows the best-fit linear relation using the errors in both
  coordinates, while the dashed line shows equality.  The scatter in the
  vertical direction is larger than in the horizontal direction, but the
  overall agreement is good.  The two ``Virgo'' galaxies with
  $\mM\approx31.8$ are actually members of the background \wprime\ group.
  The galaxy with the lowest distance is NGC\,4697. \\
\label{fig:sbfcomp}}
\end{figure}

Comparing individual galaxies, we find that our distances agree with those of
Ton01 to within the combined measurement error.  The rms scatter of the
differences is 0.24~mag, which is just slightly smaller than the quadrature
sum of the average measurement errors.  The reduced $\chi^2$ value of the
differences for these 23 galaxies is $\chi^2_N=0.93$.   There are also 26
Virgo galaxies in common between ACSVCS and Ton01; the comparison of these was
presented in \vcsxiii. We now add NGC\,4697 as well.
Figure~\ref{fig:sbfcomp} compares the full set of 50 galaxies in common between
our combined ACS sample and Ton01.  These span a factor of nearly~2 in distance,
and the reduced $\chi^2$ of the differences is $\chi^2_N=1.01$.
Fitting a linear relation with errors in both coordinates, we find
\begin{eqnarray}
(m{-}M)_{\rm Ton01} \;=\; (31.50&\pm&0.03) \;+\, \nonumber \\
                          (0.96&\pm&0.10)\,[(m{-}M)_{\rm ACS} - 31.50] \,.
\label{eq:toncomp}
\end{eqnarray}
Thus, the surveys agree closely in the mean, and the slope is consistent
with unity.  The largest Fornax outlier is FCC177 (NGC\,1380A),
which we find to be at $\mM = 31.51$, while Ton01 find $\mM =
31.00\pm0.29$, a difference of less than 2$\,\sigma$.
We note that ACS SBF measurements in the F814W bandpass (Cantiello \etal\
2007a; Barber DeGraaff \etal\ 2007) also have high precision similar to our
measurements and generally agree with Ton01, but occasionally reveal
anomalous distances in the ground-based survey.

In addition to comparing the distances, since SBF studies require very
accurate color data (or other spectral information suitable for calibration),
we can compare our measured \gz\ to \vi\ from Ton01.  One complication is that
galaxies have color gradients and the regions of the galaxies measured in the
two surveys are generally different.  The present survey analyzed
regions of the galaxies at smaller radii, due to the smaller field of view,
higher resolution, and finer pixel scale.  To account for this effect in a
similar comparison, Blakeslee \etal\ (2001a) derived an approximate
transformation from the tabulated Ton01 \vi\ values to the colors at a fixed
radius of 10\arcsec, which is within the range of radii measured here.  As a
first-order correction, we apply the same transformation, which we quote here
to correct a typo in the original work:
%\begin{equation}
$(V{-}I)_{10^{\prime\prime}} \,=\, -0.046 + 1.05\,(V{-}I)_{\rm Ton01} \, .$
%\end{equation}
The effect of this is to make the \vi\ colors slightly redder.

\begin{figure}\epsscale{1.05}
\plotone{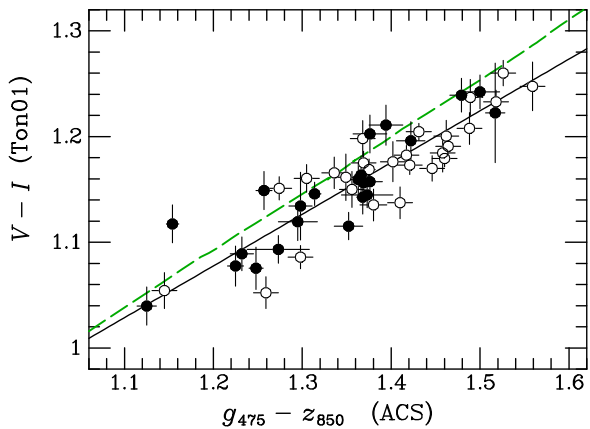}
\caption{Comparison of the ground-based \vi\ colors (Tonry \etal\ 2001)
  with our ACS \gz\ colors for the 23 Fornax (filled circles) and 26
  Virgo (open circles) galaxies in common between these two surveys.
  The \vi\ colors have been corrected according to the linear transformation
  derived by Blakeslee \etal\ (2001a).  The solid line shows the
  best-fit relation  using errors in both coordinates:
  $(V{-}I) \approx 1.15 + 0.5\,[(g{-}z)-1.35]$.
  For comparison, the dashed green line shows the locus of Bruzual \&
  Charlot (2003) solar metallicity models with ages in the range
  $\sim\,$1.5 Gyr to 15~Gyr.
  The mean \gz\ color of this matched sample is significantly redder
  than the overall sample, since the ground-based survey targeted mainly
  giant galaxies. \\
\label{fig:colorcomp}}
\end{figure}

Figure~\ref{fig:colorcomp} presents the comparison of our \gz\ values with the
transformed Ton01 colors.  Fitting a linear relation with errors in both
coordinates gives
\begin{eqnarray}
\vi \;=\; (1.151\pm0.003&\pm&0.02) \; + \nonumber \\
          (0.49&\pm&0.04)\,[(\g{-}\z)-1.35] \,, \hfil
\label{eq:vigz}
\end{eqnarray}
where the second error~bar on the zero~point is the approximate systematic
uncertainty in the tie of the ground-based data to Landolt (1992) standards
and the absolute calibration of the Landolt  \vi\ colors to a Vega-based
system.  We note that Tonry \etal\ (1997) found a zero-point scatter in \vi\
among different observing runs of about 0.02~mag.  The run-to-run offsets were
corrected by extensive intercomparisons, but the overall zero point for that
survey likely had a similar uncertainty.  The RMS scatter in the fitted relation
is 0.024~mag, which is larger than the combined observational error and
suggests an additional scatter of $\sim\,$0.015 mag due to stellar
population and aperture effects. 
For comparison, we show the locus of solar-metallicity
Bruzual \& Charlot (2003) models of varying ages.  At a given \gz, these are
redder on average by about 0.02~mag in \vi, or bluer in \gz\ by 0.04 mag
at a given \vi.  This may result from uncorrected mean color gradients,
but the agreement is reasonable given the likely uncertainties in
observational and model color zero~points.

%%%%%%%%%%%%%%%%%%%%%%%%%%%%%%%%%%%%%%%%%%%%%%%%%%%%%%%%%%%%%%%%%%
%%%%%%%%%%%%   a few other optical SBF studies     %%%%%%%%%%%%%%%
%%%%%%%%%%%%%%%%%%%%%%%%%%%%%%%%%%%%%%%%%%%%%%%%%%%%%%%%%%%%%%%%%%

There have been a few other optical SBF studies in Fornax, but the overlap
with the present sample is much less.  Blakeslee \etal\ (2001b) measured and
calibrated $V$-band SBF magnitudes for the five brightest galaxies in the
ACSFCS sample.  Although the mean distance was lower by about 0.15 mag due to
a different choice of zero-point calibration, the galaxies were consistent
with being all at a common distance within the errors, the same as we find here.
Jerjen (2003) and Dunn \& Jerjen (2006) measured
$R$-band SBF distances for a total of 18 dwarf FCC galaxies.  However, only
three of these galaxies overlap with our sample; the rest are fainter than our
magnitude limit.  For the three galaxies in common, the $R$-band SBF distances
are larger by $0.21\pm0.05$ mag.  However, this is probably due to small
number statistics, as the overall mean distance modulus from Dunn \& Jerjen
is 31.50 mag, very similar to ours. 
The rms dispersion of their distance moduli is 0.21~mag, the same
as in Ton01 and dominated by measurement errors.  Finally, we note that 
Mieske et al.\ (2006) report $I$-band SBF measurements for 21 FCC dwarfs, mainly 
to evaluate the SBF scatter at blue colors, but all
are quite faint and below our magnitude limit.

\subsection{Near-IR SBF in Fornax}

In addition to the optical surveys, there have been two near-IR SBF studies
including significant numbers of Fornax galaxies.  In the first of these, Liu
\etal\ (2002) measured ground-based $K_s$ SBF magnitudes for 19 Fornax
galaxies in $\sim\,$1\arcsec\ seeing with a 256$^2$~pix IR array.  There are
17 galaxies in common with our sample.  The $\overline K_s$ distances
show a large scatter when compared to ours, in excess of 0.4~mag, giving
a very poor $\chi^2_N$.  This may reflect anomalous behavior of SBF in the $K_s$
bandpass, with interesting stellar population implications.  However, we
believe it more likely reflects the difficulty in obtaining accurate
ground-based $K$-band SBF measurements at a distance of 20 Mpc with small a
detector, mediocre seeing and a very bright sky background.

\begin{figure}\epsscale{1.05}
\plotone{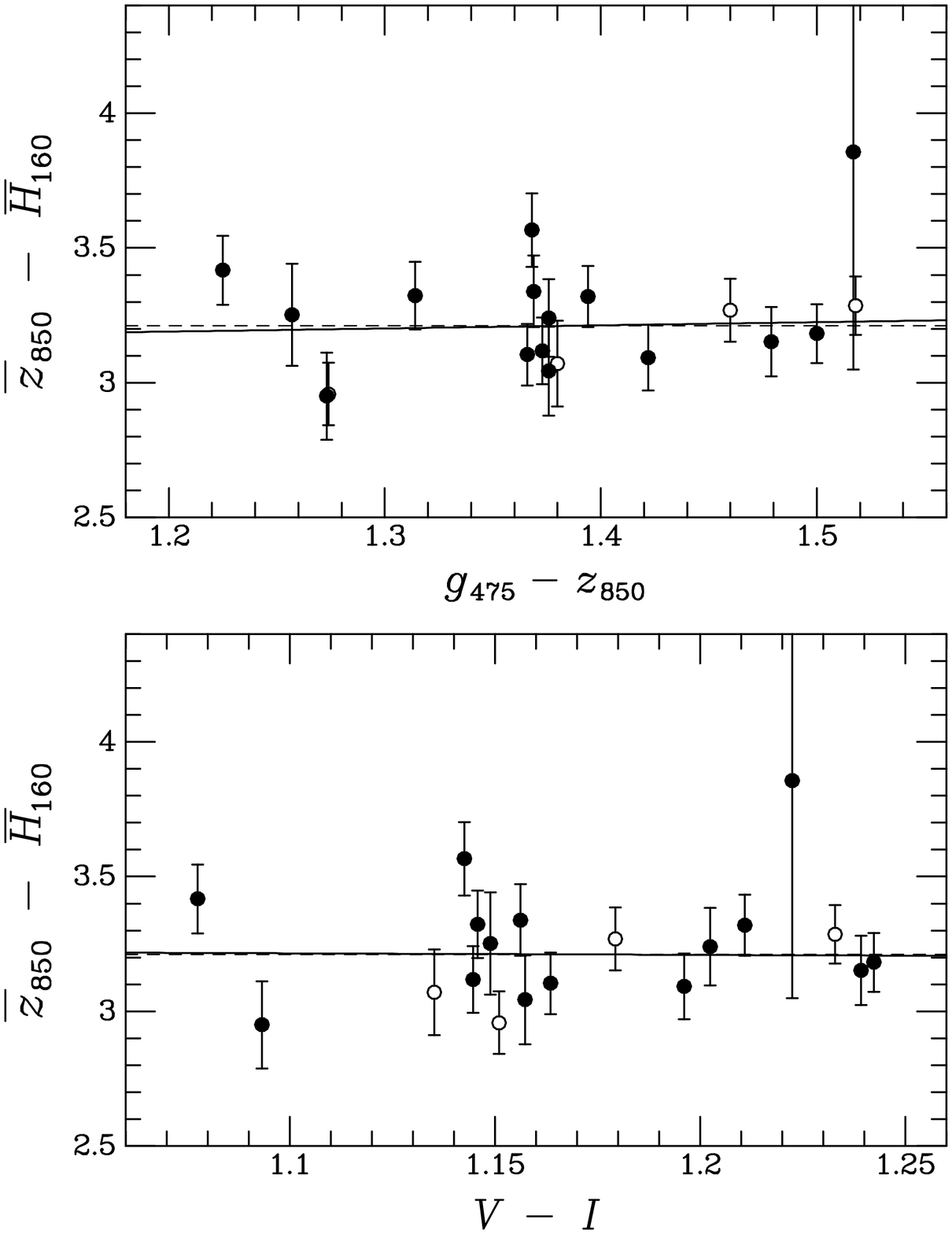}
\caption{The  $\zbHb$ ``fluctuation color'' is plotted against \gz\ and \vi\ 
  photometric colors for galaxies having both ACS \zbar\ and NICMOS
  \Hbar\ measurements (the latter from Jensen \etal\ 2003).  
  Four Virgo galaxies (open circles) and 15 Fornax galaxies (filled
  circles) have both \zbar\ and \Hbar\ data.
  The dashed line in each panel shows the weighted mean, while the solid
  line is the best-fitting linear relation.  In both cases the slope is
  indistinguishable from zero, and the scatter is consistent with measurement
  errors.  Thus, within observational limits, \zbar\ and \Hbar\ have the
  same dependence on stellar population. 
\label{fig:zbar_Hbar}}
\end{figure}

In another study, Jensen \etal\ (2003) measured SBF magnitudes for 19 Fornax and 4
Virgo galaxies (as part of a larger archival sample) in the NICMOS F160W
bandpass ($H_{160}$).  Their sample included the bulges of several spiral
galaxies, and the main goal of this work was to calibrate the behavior \Hbar\
as a function of galaxy color.  The ACSFCS and ACSVCS samples combined have 19
galaxies in common with Jensen \etal\ (2003), and we find it most interesting
to present a direct comparison of the SBF magnitudes.
Figure~\ref{fig:zbar_Hbar} plots the ``fluctuation color'' $\zbar{-}\Hbar$ as
a function of broadband \gz\ and \vi\ colors for these galaxies.  The rms
scatter of 0.21~mag in $\zbar{-}\Hbar$ is consistent with measurement errors,
and there is no evidence for a dependence on the broadband color.  This
implies that $\Hbar$ will be extremely useful for future SBF studies when a
modern, wide-area, space-based near-IR camera such as WFC3/IR becomes
available.  One cautionary note is that most of the galaxies in
Figure~\ref{fig:zbar_Hbar} are fairly red, and the scatter in \Hbar\ shows
evidence of increasing at bluer colors (Jensen \etal\ 2003). However, the
prospects for \Hbar\ with WFC3/IR appear promising.

\begin{figure}\epsscale{1.05}
\plotone{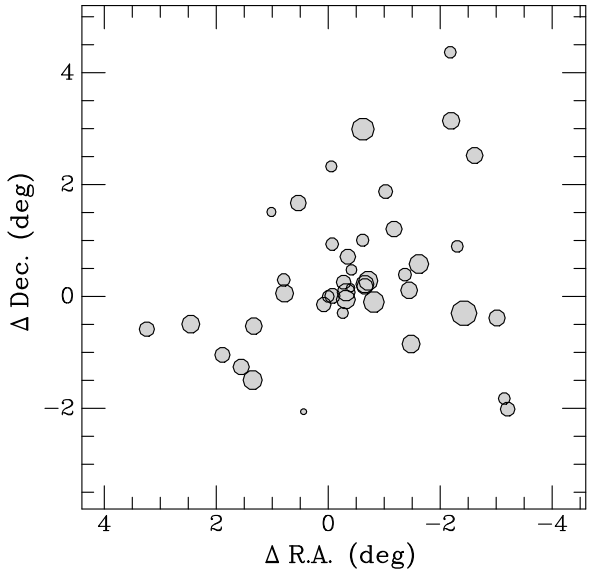}
\caption{%
Spatial distribution of the ACS Fornax cluster galaxies.  The offsets
in Right Ascension and Declination are computed with respect to the cD
galaxy NGC\,1399.  This plot is similar to Figure~2 of Jordan \etal\ (2007), 
except here the point size is scaled inversely with the square
of the SBF distance.  The range in distance (point size) appears to
increase with radius from the cluster center; for instance, the most
distant galaxy FCC249 is also the farthest south, and the two nearest
galaxies with $d<19$ Mpc are both at radii $r>2^\circ$, well beyond
the cluster core.  However, there is
no obvious trend of distance with position along any preferred axis.
\label{fig:sky}}
\end{figure}

\subsection{Structure of Fornax}
\label{ssec:struct}

Figure~\ref{fig:sky} shows the spatial distribution of the 43 ACSFCS galaxies
plotted with their symbol sizes scaled inversely with SBF distance.  The range
of distances appears to increase slightly for galaxies farther away from the
cluster center. This is not unexpected, since the highest density of galaxies
will be near the center of the cluster, and the area is small, so there will
be a lower incidence of galaxies projected from larger radii. 
The 21 galaxies at radii $r<1\fdg5$ have an rms distance scatter of 0.080
mag, while the 22 at larger radii have an rms scatter of 0.105 mag.  However,
the significance of this difference is only 1.1$\,\sigma$.  Overall, the
cluster appears compact with no obvious trends of distance with position.
The two galaxies in our sample outside the FCC survey area, NGC\,1340 and
IC\,2006, both have distances near the mean of the full sample.~ 

\begin{figure}\epsscale{1.1}
\plotone{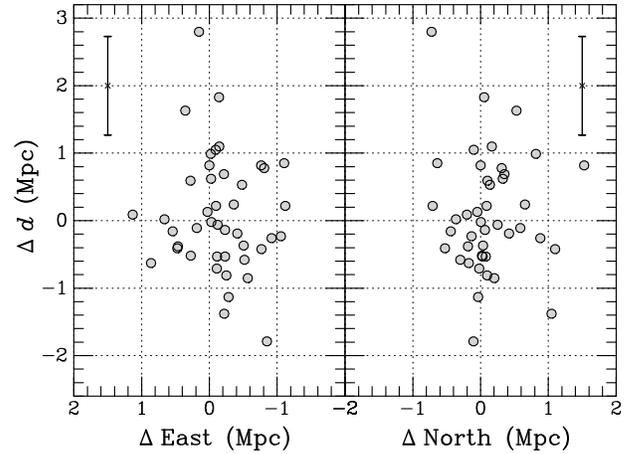}
\caption{%
Galaxy distance (with respect to the mean of 20 Mpc) is plotted
versus physical offset in Mpc east-west (left panel) and north-south (right
panel) with respect to NGC\,1399.  Although different ranges are
plotted along the horizontal and vertical axes, the scale is the same
in both directions.    The cluster appears fairly symmetric in the
distance-RA plane (left) and somewhat narrower in the Declination
direction (right). However, this is mainly because of distance
measurement errors; the median error in $\Delta d$ is shown in each
panel.  There is a bias towards the cluster
appearing elongated along the line of sight due to distance errors
and because galaxies more than
about $\pm1.5$ Mpc from the cluster mean would not be included in the
FCC catalogue if the offset were in the plane of the sky rather than
along the line of sight.
\label{fig:dist_xy}}
\end{figure}

% This is shown further
The compact structure is further illustrated
 in  Figure~\ref{fig:dist_xy}, which plots offset in
line-of-sight distance versus projected physical distance in the RA and Dec
directions from the
central galaxy.  As noted in the caption, there is a bias for the cluster to
appear elongated along the line of sight in this figure, both because of
distance measurement errors and selection effects in an area-limited survey.
However, the central Mpc looks fairly symmetric.
The rms dispersions in projected distance east-west and north-south of
the cluster center are $0.52\pm0.06$~Mpc and $0.47\pm0.06$ Mpc, respectively,
where the errors are estimated from bootstrap resampling.  If we instead use robust
biweight estimates, these become $0.52\pm0.07$ Mpc and 
$0.44\pm0.08$ Mpc, respectively, not sensibly different from the rms values.

To estimate the true rms linear depth, we correct for the mean measurement error
$\sigerr=0.047$ mag and the SBF method intrinsic scatter $\sigcos =
0.06\pm0.01$ mag, to obtain $\sigma_d = 0.49^{+0.11}_{-0.15}$~Mpc, where the
quoted error bars are the 1-$\sigma$ uncertainties on the rms depth $\sigma_d$.
The $\pm\,$2-$\sigma_d$ distance depth of Fornax is then $2.0^{+0.4}_{-0.6}$~Mpc.
% bootstrap error
However, this estimate is somewhat circular because we used the spatial
distribution on the sky to constrain the value of \sigcos\ in the $\chi^2$
analysis of \S\,\ref{sec:cal}.  For example, if we set $\sigcos=0$, then the
depth estimate increases to $0.74^{+0.15}_{-0.20}$~Mpc.  This therefore sets a
firm upper limit of $\sigma_d < 0.9$~Mpc on the rms depth of Fornax.  Conversely,
we can also force the depth to zero by setting $\sigcos\approx0.08$~mag, which
sets a limit on the intrinsic scatter of the \z-band SBF method to be
$\sigcos\lta0.08$ mag.  The compact nature of Fornax allows us to place firmer
limits on \sigcos\ than did the Virgo cluster, where the rms depth in
magnitudes is larger than the upper limit on~$\sigcos$.

\begin{figure}\epsscale{1.05}
\plotone{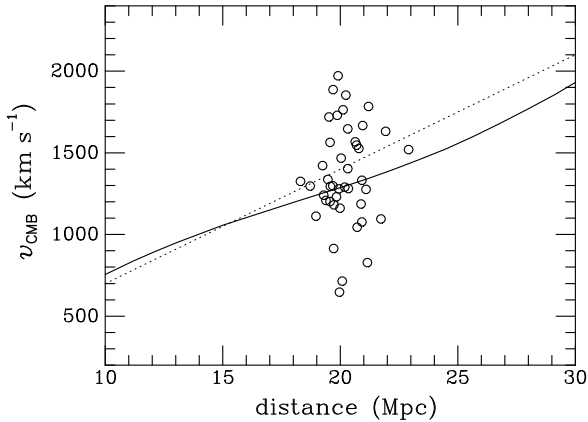}
\caption{%
Hubble diagram for our 43 ACSFCS galaxies.  The solid curve shows the
 velocity-distance relation along the Fornax line of sight as predicted by the 
\textit{IRAS} galaxy density field (Blakeslee \etal\ 2002).  The
dotted line shows a pure Hubble flow for $H_0 = 70$ \kmsMpc.  
As expected for a spatially compact, virialized structure, there is
no relation between distance and velocity for this 
sample of early-type Fornax cluster galaxies.\\
\label{fig:hubble}}
\end{figure}

Figure~\ref{fig:hubble} presents the velocity-distance relation for our
sample of Fornax galaxies.  In contrast to the results for Virgo from
\vcsxiii, we do not find evidence for non-virial motions in our sample.  Both
the velocities and distances have much smaller ranges in  Fornax than in
Virgo, and we find no correlations between them in our limited sample.  The
cluster appears to be well virialized.  Unlike Virgo or Centaurus, the two
massive attractors that dominate the local velocity field (with likely a
residual dipole due to more distant mass concentrations), Fornax has no
surrounding supercluster.  It was not detected as an attractor in the velocity
field analysis by Tonry \etal\ (2000); their model $\chi^2$ was not
significantly improved by adding the additional parameters for a velocity
attractor at the location of Fornax.  Instead, that work simply assigned the
Fornax cluster galaxies an increased velocity dispersion equal to about twice
the background thermal dispersion, and about half that of Virgo.

However, the signature of the mass of Fornax is present in the
velocity-distance curve inferred from the \textit{IRAS} galaxy density field,
and we show the relation for the Fornax line of sight from Blakeslee \etal\
(2002), which used $H_0 = 73$ \kmsMpc.  For comparison, the dotted line shows
an unperturbed Hubble flow with $H_0 = 70$ \kmsMpc.
While it would be unwarranted to present a value for $H_0$ from a single nearby
cluster, even with a highly accurate distance, the consistency of the mean
distance and velocity within the accepted range of $H_0$ is reassuring.

We note that Dunn \& Jerjen (2006) combined their sample of dwarf galaxy SBF
distances with those of Jerjen (2003) and Ton01 and proposed that a composite
sample of 29 early-type galaxies within $2^\circ$ of NGC\,1399 showed some
evidence for the \textsf{S}-shaped signature of galaxy infall.  This feature
is not evident in the Ton01 data by itself.  Further, Dunn \& Jerjen noted
that the measurement accuracy was ``not sufficient to establish the reality of
the effect,'' and concluded that there was ``no evidence for elongation along
the cluster line of sight'' even in the composite sample.  However, their
Figure~7 remains suggestive.  It may be that the fainter Fornax dwarfs which
constituted the Dunn \& Jerjen samples (all but three of which are fainter
than our sample magnitude limit) are preferentially infalling into the
cluster.  This is an intriguing possibility that deserves following up with
high-precision \hst\ SBF distances for a fainter sample of Fornax dwarfs.

\subsection{Fluctuation Count \Nbar\ and an Alternate SBF Calibration} 
\label{ssec:nbar}

The absolute SBF magnitude \Mbar\ in a given bandpass has a corresponding
``fluctuation luminosity'' \Lbar, which is equal to the luminosity-weighted
mean luminosity of the stellar population in a galaxy (or star cluster, etc).
The ratio $\Ltot/\Lbar$ gives the total galaxy luminosity in units of \Lbar,
which depends on the stellar population and generally gets fainter for redder
galaxies, as shown by the calibrations derived here and elsewhere.  Viewed
another way, this ratio represents the number of stars of luminosity
\Lbar\ needed to constitute the luminosity of the galaxy.
Motivated by these considerations,
Tonry \etal\ (2001) introduced the distance-independent ``fluctuation~count''
\Nbar, defined in a given bandpass as 
\begin{equation}
\Nbar \,=\, \mbar - m_{\rm tot} \,=\,
    +2.5\,\log\left[L_{\rm tot} \over \lbar\right] \,.
\label{eq:nbardef}
\end{equation}
Besides being distance independent, \Nbar\ is also independent of Galactic
extinction, or even photometric calibration error if \mbar\ and \mtot\ are
determined from the same imaging material.
However, it correlates well with galaxy color, a consequence of the
mass-metallicity relation for early-type galaxies, as discussed by
Blakeslee \etal\ (2001b).   Tonry \etal\ (2001) found that
the scatter in the predicted value of \vi\ at a given \NIbar\ was about
0.04~mag, based on a linear fit over the range $17\lta\NIbar\lta23$.

\begin{figure}\epsscale{1.05}
\plotone{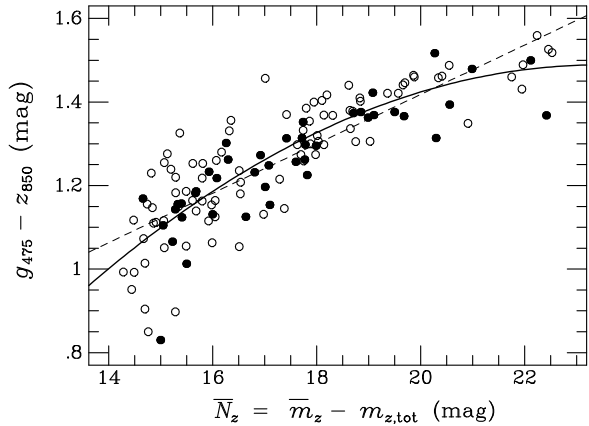}
\caption{%
Galaxy \gz\ color is plotted as a function of the ``fluctuation 
count'' $\Nzbar$, the difference between the SBF magnitude \zbar\
and the total $z$ magnitude of the galaxy.  
Open and filled circles 
represent Virgo and Fornax galaxies, respectively.  Note that \Nzbar,
like \gz, is independent of galaxy distance and can therefore be used
as another means of calibrating the SBF distances, but unlike color,
it is also independent of Galactic extinction.  Tonry \etal\ (2001)
showed that \NIbar\ scaled approximately linearly with \vi\ over the range
$16.5<\NIbar<22.5$.  Here, we extend the range by another two magnitudes;
the resulting relation exhibits curvature at the blue/low-luminosity end.
The dashed line shows the best-fit linear relation, while the solid
curve is a quadratic fit. \\[2pt]
\label{fig:gzNbar}}
\end{figure}

Figure~\ref{fig:gzNbar} shows the relation between \Nzbar\ and \gz\ for the
sample of 133 galaxies with SBF measurements from our Virgo and Fornax surveys.
The total \z-band magnitudes are from \cote\ \etal\ (2008, in preparation; see also
\vcsvi). These data
cover a much larger range in luminosity $14.5\lta\Nzbar\lta22.5$ and 
stellar population.  The range in \gz\ is more than twice the \vi\ range
used to the define the \Nbar\ relation by Tonry \etal\ (2001), since that
survey concentrated mainly on bright ellipticals and S0s.  Our sample contains
these galaxies as well, but is dominated by dwarfs, and we find that the relation
between \gz\ and \Nzbar\ is nonlinear for our full sample.
It may be approximated by the simple least-squares linear and quadratic fits
\begin{eqnarray}
\gz &\,=\,&   1.300 \,+\, 0.059\,(\Nzbar-18) \;, \\
\gz &\,=\,&   1.327 \,+\, 0.060\,(\Nzbar-18) \,-\, 0.0055\,(\Nzbar-18)^2\,,~~~~~~~
\end{eqnarray}
which are plotted in Figure~\ref{fig:gzNbar} and
yield rms scatters in \gz\ of 0.083~and 0.079~mag, respectively.  The
scatter also increases at low \Nzbar\ in a manner similar to that of
standard galaxy color-magnitude diagrams.

The correlation of the distance-independent \Nzbar\ with color means
that it can be used to define an alternative SBF calibration, which may
be useful especially in cases of large or uncertain Galactic extinction.
Surprisingly, Blakeslee \etal\ (2002) found that the Tonry \etal\ (2001)
SBF distances showed 25\% less scatter when recalibrated using
\NIbar\ instead of \vi\ in a comparison to the peculiar velocity
predictions from the \textit{IRAS} galaxy density field.  This was
explained by the greatly reduced sensitivity to photometric measurement
errors (a small error in $V{-}I$ would be amplified by a factor of 4.5 
from the $I$-band SBF calibration) and Galactic extinction, an important
consideration for an all-sky survey.  In addition, the ground-based
SBF galaxy sample was comparatively homogeneous in terms of galaxy type,
with few if any dwarfs beyond the Local Group.

\begin{figure}\epsscale{1.04} 
\plotone{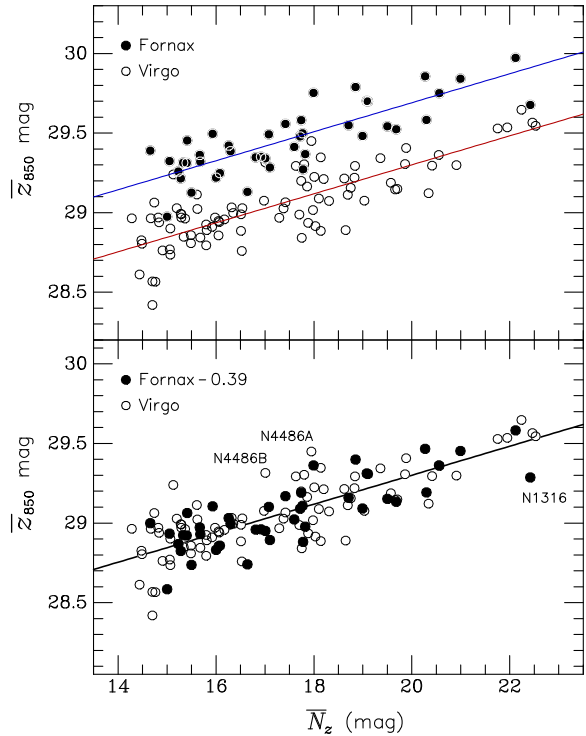}
\caption{\small
  An alternative SBF calibration, similar to Figure~\ref{fig:finalcal}, but 
  \zbar\ is here plotted as a function of~\Nzbar.
  A~linear fit provides an excellent description for this relation,
  although the estimated internal scatter is about 45\%
  larger than the polynomial calibration against \gz\ color.  The
  lower panel shows the two samples shifted together by subtracting
  the best-fit relative modulus $\mMfv=0.39\pm0.03$
  from the Fornax galaxy distance moduli.
  The very luminous, blue, post-merger  galaxy FCC21 (NGC1316,
  labeled) is a prominent outlier and, because of its position at the
  bright end of the relation, has an undue influence on the
  slope; it has been omitted from the fit.  The two M87 companion
  VCC1327 (NGC\,4486A) and VCC1297 (NGC\,4486B), which may be expected
  to deviate in the other sense from tidal stripping are also labeled.  \vspace{6pt}
\label{fig:nbarcal}}
\end{figure}

To test the usefulness of \Nbar\ for distance measurements, we repeated
the same $\chi^2$ analysis as in \S\,\ref{sec:cal} above, except
substituting \Nbar\ for \gz.  We find that the SBF calibration based on
\Nzbar\ is well described by a linear relation.
Figure~\ref{fig:nbarcal} shows the resulting \Nzbar\ SBF calibration
equivalent to Figure~\ref{fig:finalcal} for \gz.   In this case, the
calibration no longer has a pure basis
in stellar population properties, but rather brings
in a scaling relation between galaxy luminosity and color.  Therefore,
any very luminous blue galaxy with recent star formation, or small red galaxy,
perhaps tidally stripped like M32, should be expected to
deviate from the relation.  A particularly conspicuous outlier in this
regard is the merger remnant NGC\,1316 (FCC21), the brightest galaxy in Fornax.
Its position near the extremum of the relation gives it a large
influence on the slope; since it is a known outlier, we have removed it
from the fit.  However, it has little effect on the final relative
distance modulus or internal scatter estimate, both of which change by less than
0.01~mag when it is included.  We also label the two galaxies NGC\,4486A
(VCC1327) and NGC\,4486B (VCC1297), which are close companions of M87
and have likely undergone tidal stripping.  They would be expected to lie
above the mean relation, and we find that this is indeed the case.

We use the same allowance for cluster depths as in the preceding section
and find that a relative distance modulus of 
$\mMfv = 0.39\pm0.03$ minimizes the value of $\chi^2$.  The best-fit
relation is given by
\begin{equation}
\overline M_z \;=\; -1.98 \,+\, 0.089\,(\Nzbar - 18) \,.
\label{eq:Ncal}
\end{equation}
The magnitude zero-point differs from that in equation~(\ref{eq:cal})
because $\Nbar=18$ does not correspond precisely  to $\gz=1.3$.
Note that we can use the definition of \Nzbar\ to rewrite this equation for
the distance modulus:
\begin{equation}
 \mM  \;\approx\;  3.58  \,+\,  0.91\,\zbar\  \,+\,  0.09\,\ztot\ 
\label{eq:mMNbarl}
\end{equation}
where \ztot\ is the total apparent \z\ magnitude of the galaxy.  This shows
that distances calculated from \Nzbar\ are $\sim\,$9\% less sensitive to
the SBF \zbar\ measurement errors and fairly insensitive to errors in the
total magnitude.  However, recall that this is all based on an empirical
scaling relation.  Ton01 cautioned that it may have systematic
environmental or type dependencies; both these effects are
seen in Virgo for the closely related color-magnitude relation (Lisker
\etal\ 2008).

In order to have $\chi^2_N=1.0$, the \Nzbar\ calibration in
equation\,(\ref{eq:Ncal}) must have a cosmic scatter ${\sigcos}(\Nzbar) \approx
0.10$~mag, which is $\gta\,$60\% larger than the intrinsic scatter for the calibration
based on \gz.
Thus, we find that the SBF distances are more accurate when calibrated against
color, in contrast to the ground-based data.
This is likely because of the
much higher precision of our measurements, which have errors smaller than the
intrinsic scatter in the calibration relations, unlike the ground-based SBF
data, where the measurement errors dominated.  In addition, Galactic
extinction is not a big issue here, since we are only dealing with two
sightlines, both of which have minimal extinction.  Further, our color
calibration implies $\delta \zbar/\delta\gz\approx1.5$, whereas the $I$-band
calibration has slope $\delta \Ibar/\delta\vi \approx 4.5$.  Therefore, the
ground-based $I$-band SBF distances are a factor of three more sensitive than
ours to photometric errors in the color measurement.  All of these factors 
contribute to the high precision of our \gz-calibrated SBF distances; we prefer
to adopt the relative distance modulus from that analysis, given its firm
grounding in relatively well understood stellar population properties.

\subsection{A Final Note on Virgo}

Before concluding, we wish to note that our revised calibration and
retabulated distances have no effect on the conclusions of
\vcsxiii\ regarding the structure of Virgo.  The rms difference in the
distance moduli between our recalibrated Virgo distances and those used for
the analysis of \vcsxiii\ is only 0.02 mag.  Only three galaxies (the reddest
ones) have moduli differing by more than 0.05~mag, in the sense of the
recalibrated values being lower, and the maximum difference is $0.10$~mag.
None of these changes affects our earlier conclusions.  However, we have
presented the revised values in Table~\ref{tab:virgo} in order to make the 
most homogeneous set of Virgo and Fornax galaxy distances, based on our
final calibration for this project, readily available.

\section{Summary and Conclusions}
\label{sec:summary}
%   - what we did
%   - what we found
%   - what comes next

We have presented new SBF magnitude and color measurements for 43 galaxies
observed in the ACS Fornax Cluster Survey.  We first used these data to fit the
variation of the apparent SBF magnitude \zbar\ with galaxy \gz\ color based on
the ACSFCS data alone.  The relation is nonlinear, as found previously in
\vcsxiii.  For a quadratic fit, the observed scatter in the Fornax \zbar\
versus \gz\ relation is 0.092~mag.  This contains contributions from
measurement error ($\sim0.047$ mag), true distance variations within Fornax
($\sim0.053$ mag), and the cosmic scatter \sigcos\ in \Mbarz\ at a given \gz\
due to stellar population effects ($\sim0.060$~mag).  This is the first large
SBF study in which measurement errors and depth effects are each smaller than
the intrinsic scatter, affording us the opportunity for interesting constraints
on \sigcos.
% e.g., the magnitude scatter in Virgo due to cluster depth is about 40\% larger.
We determined $\sigcos = 0.06\pm0.01$, with a firm upper limit $\sigcos<0.08$~mag,
as this would require the cluster to have zero depth.  This result holds
for the combined Virgo+Fornax sample of galaxies with
$\gz>1.02$~mag.  For the small number of bluer galaxies, mainly in Virgo,
we estimate $\sigcos\approx0.13$~mag.

The combination of our Fornax SBF sample with that of Virgo from \vcsxiii\
represents the largest homogeneous set of SBF measurements available, with 
nearly identical observing and data analysis procedures. 
The few modifications to the procedures were done in order to make the
analysis more robust and automatic, and we have described them in detail.
However, the Virgo galaxy selection extended to fainter luminosities,
thus including a more significant blue tail of galaxies.  Despite this, the
overall means of the galaxy color distributions are very similar.  We used 
$\chi^2$ minimization to fit a single polynomial calibration for $\Mbarz$ in
terms of \gz\ for both Virgo and Fornax galaxies, assuming a
range of relative distance moduli for the two clusters.
Thus, we derived the following revised calibration, valid for $0.8<\gz<1.6$:
\begin{equation}
\overline M_z \;=\; (-2.04\pm0.15) \,+\, 1.41\,x \,+\, 2.60\,x^2 \,+\,3.72\,x^3 \,,
\label{eq:finalcal}
\end{equation}
where $x \,\equiv \,\gz - 1.3$.  
The error on the \Mbarz\ zero~point comes
from the total uncertainty in the Cepheid distance scale and the tie between
spiral and early-type galaxy distances.  The intrinsic scatter about 
equation~(\ref{eq:finalcal}) is again 0.06~mag.

We have tabulated recalibrated distances for a total of 134 early-type galaxies
from the two surveys and NGC\,4697 in the Virgo Southern Extension.
The calibration procedure determined a best-fit relative
Fornax-Virgo distance modulus $\mMfv = 0.42\pm0.02$ mag, where the error bar
(1$\,\sigma$) comes from the $\chi^2$ analysis and was verified by bootstrap
resampling.  This result is robust, regardless of whether or not the blue tail
of galaxies is included in the fit: the full sample requires the cubic
polynomial calibration above, whereas a quadratic suffices for $\gz>1.02$~mag,
but the best-fit relative distance is the same. %
After considering all possible systematic effects, we estimate the total
uncertainty on \mMfv\ is $\pm0.03$~mag. For an adopted Virgo distance
modulus of $(m{-}M)_{{\rm Vir}} = 31.09\pm0.15$~mag, the resulting Fornax
modulus is $(m{-}M)_{{\rm For}} = 31.51\pm0.03\pm0.15$ mag, or a distance of
$d_F = 20.0\pm0.3\pm1.4$~Mpc.  The physical distance between the centers of
Virgo (M87) and Fornax (NGC\,1399) is then $33.4\pm0.5$~Mpc.

%%% structure of Fornax

Correcting for measurement error and internal/cosmic scatter in the method, we estimate
the rms line-of-sight depth of Fornax cluster to be
$\sigma_d = 0.49^{+0.11}_{-0.15}$~Mpc.
The total back-to-front depth of Fornax (i.e., $\pm2\sigma_d$) is
therefore about $2.0^{+0.4}_{-0.6}$~Mpc, meaning that the cluster galaxies span a
distance range of 19--21~Mpc.  
However, this estimate depends on our value of $\sigcos=0.06\pm0.01$ mag for
the cosmic scatter.  If we had unrealistically assumed $\sigcos\equiv0$ (i.e.,
SBF is a perfect distance indicator with no internal scatter), then the rms
depth estimate would increase to $0.74^{+0.15}_{-0.20}$~Mpc; this provides an
upper limit to the true rms depth of Fornax.  We no not find any
evidence for trends in the galaxy distances along any preferred direction on
the sky, nor do we find evidence for non-virial motions.  In particular, there
is no sign of ongoing cluster infall among the galaxies in our sample.
It would be interesting to see if this also holds true for faint cluster dwarfs
below our magnitude limit and later type galaxies.

%%% lit comparison

We compared our ACS Virgo and Fornax SBF measurements to the ground-based
values of Tonry et al.\ (2001) for 50 galaxies in common. Our measurement
errors are a factor of three smaller than those from the ground-based survey.
Overall, the distances agree within the errors without any adjustments.  
We also compared our \zbar\
measurements to the near-IR \Hbar\ data from Jensen \etal\ (2003) for 19
galaxies in common.  The resulting SBF color $\zbar{-}\Hbar$ exhibits a scatter
consistent with measurement error and no dependence on the integrated galaxy
color, although the sample is limited to fairly red galaxies.  This suggests
that $\Hbar$ is capable of providing large samples of excellent SBF distances
with less sensitivity to dust extinction, an important consideration for
all-sky surveys. 

%%% Nbar

The distance-independent fluctuation count $\Nbar = \mbar - \mtot$ scales
logarithmically with the number of stars in a galaxy, and is therefore related
to the total mass.  It correlates with galaxy color because of the
mass-metallicity scaling relation.  We presented linear and quadratic fitting functions
for the dependence of \Nbar\ on \gz\ in the combined cluster sample.  We
then repeated the \Mbarz\ calibration fit, but with \Nbarz\ as the
distance-independent parameter instead of \gz.
The resulting calibration is linear, but bright blue galaxies with
recent star formation and small red galaxies that have undergone tidal
stripping are expected, and observed, to be outliers.  Further, the intrinsic
scatter in the \Mbarz--\Nbarz\ relation is 0.10~mag, or 2/3
larger than for the standard \Mbarz--\gz\ calibration, but the
best-fit relative distance modulus of $0.39\pm0.03$ mag 
agrees with the result from the standard calibration.  However, 
we prefer the relative distance from the \gz\ analysis because
of its smaller scatter and pure basis in stellar population properties.

%%% The Undiscovered Country
%
% ACS has shown the potential of the SBF method in mapping the structure of
% the Virgo cluster and determining F-V relative distance to high precision 
% Additional SBF HST studies ongoing (shapley)
% Deeper ground-based surveys may improve on sky coverage of the ground-based
% I-band survey, but with uniformly better quality
%
% H-band looks great, if have WFC3
% ground Wide-field near-IR detectors may clean up

The ACS Virgo and Fornax cluster surveys have shown the power of the SBF
method when coupled with an instrument such as ACS.
In \vcsxiii, we successfully mapped the 3-D structure of the Virgo cluster,
and here we have determined a precise value for the relative distance of Fornax
with respect to Virgo.  Additional SBF studies with ACS data are nearing
completion and will provide new insight into the sources of the Local Group
motion [see Blakeslee \& Barber DeGraaff (2008) for an example of these data].  
Future all-sky optical surveys reaching faint limits from the
ground should provide large amounts of uniform quality imaging data for SBF
measurements.  At longer wavelengths, the comparison of ACS and NICMOS SBF
data hints at the superb potential for near-IR SBF studies once a
modern, space-based IR camera comparable to ACS becomes available.
New wide-area IR detectors with excellent cosmetic properties now
available at many ground-based observatories present further exciting opportunities for
SBF distance and stellar population studies, as multi-band optical/IR
SBF data are relevant to such problems as AGB evolution and the UV excess in elliptical
galaxies (e.g., Buzzoni \& Gonzalez-Lopezlira 2008).  It will be interesting to
see all the directions taken by SBF research in its second twenty years.

%%%%%%%%%%%%%%%%%%%%%%%%%%%%%%%%%%%%%%%%%%%%%%%%%%%%%%%%%%%%%%%%%%%%%%%%%%

\acknowledgements

Support for programs GO-10217 and GO-9401 was provided through grants from
the Space Telescope Science Institute, which is operated by the Association
of Universities for Research in Astronomy, Inc., under NASA contract
\hbox{NAS5-26555}.  This research has made use of the NASA/IPAC
Extragalactic Database (NED) which is operated by the Jet Propulsion
Laboratory, California Institute of Technology, under contract with the
National Aeronautics and Space Administration.  We thank the anonymous
referee for thoughtful comments that helped to improve the paper.

{\it Facility:} \facility{HST (ACS/WFC)}

\clearpage

\def\tmk#1{\tablenotemark{#1}}
% table generated by textab.py !
\begin{deluxetable}{lrrrrrc}
\tabletypesize{\small}
\tablecaption{SBF Data for Fornax Galaxies\label{tab:fornax}}
\tablewidth{0pt}
\tablehead{
\colhead{Galaxy} &
\colhead{$(g{-}z)$} &
\colhead{$\mbarz$} &
\colhead{\mM} &
\colhead{$d$} &
\colhead{$B_T$} &
\colhead{Name} \\
\colhead{(1)} &
\colhead{(2)} &
\colhead{(3)} &
\colhead{(4)} &
\colhead{(5)} &
\colhead{(6)} &
\colhead{(7)}
}
\startdata
 FCC19        & 1.066 $\pm$ 0.025 & 29.258 $\pm$ 0.036 & 31.534 $\pm$ 0.074 & 20.3 $\pm$ 0.7  &  15.2  & ESO301-08 \\      
 FCC21        & 1.368 $\pm$ 0.007 & 29.676 $\pm$ 0.020 & 31.606 $\pm$ 0.065 & 21.0 $\pm$ 0.6  &   9.4  & NGC1316 \\        
 FCC26\tmk{c} & 0.830 $\pm$ 0.025 & 28.974 $\pm$ 0.055 & 31.489 $\pm$ 0.139 & 19.9 $\pm$ 1.3  &  15.0  & ESO357-25 \\      
 FCC43 	      & 1.154 $\pm$ 0.007 & 29.283 $\pm$ 0.039 & 31.485 $\pm$ 0.073 & 19.8 $\pm$ 0.7  &  13.5  & ESO358-01 \\      
 FCC47 	      & 1.298 $\pm$ 0.013 & 29.271 $\pm$ 0.040 & 31.314 $\pm$ 0.075 & 18.3 $\pm$ 0.6  &  13.3  & NGC1336 \\        
 FCC55 	      & 1.248 $\pm$ 0.008 & 29.492 $\pm$ 0.051 & 31.598 $\pm$ 0.080 & 20.9 $\pm$ 0.8  &  13.9  & ESO358-06 \\      
 FCC63 	      & 1.373 $\pm$ 0.029 & 29.548 $\pm$ 0.019 & 31.470 $\pm$ 0.083 & 19.7 $\pm$ 0.8  &  12.7  & NGC1339 \\        
 FCC83 	      & 1.363 $\pm$ 0.017 & 29.482 $\pm$ 0.020 & 31.422 $\pm$ 0.071 & 19.2 $\pm$ 0.6  &  12.3  & NGC1351 \\        
 FCC90 	      & 1.013 $\pm$ 0.047 & 29.126 $\pm$ 0.144 & 31.445 $\pm$ 0.193 & 19.5 $\pm$ 1.7  &  15.0  & \dots \\          
 FCC95 	      & 1.262 $\pm$ 0.013 & 29.385 $\pm$ 0.037 & 31.476 $\pm$ 0.073 & 19.7 $\pm$ 0.7  &  14.6  & \dots \\          
FCC100 	      & 1.105 $\pm$ 0.011 & 29.324 $\pm$ 0.048 & 31.568 $\pm$ 0.078 & 20.6 $\pm$ 0.7  &  15.5  & \dots \\          
FCC106 	      & 1.186 $\pm$ 0.017 & 29.320 $\pm$ 0.025 & 31.493 $\pm$ 0.068 & 19.9 $\pm$ 0.6  &  15.1  & \dots \\          
FCC119 	      & 1.182 $\pm$ 0.018 & 29.363 $\pm$ 0.077 & 31.539 $\pm$ 0.100 & 20.3 $\pm$ 0.9  &  15.0  & \dots \\          
FCC136 	      & 1.218 $\pm$ 0.020 & 29.248 $\pm$ 0.038 & 31.388 $\pm$ 0.075 & 18.9 $\pm$ 0.7  &  14.8  & \dots \\          
FCC143 	      & 1.273 $\pm$ 0.035 & 29.350 $\pm$ 0.041 & 31.427 $\pm$ 0.086 & 19.3 $\pm$ 0.8  &  14.3  & NGC1373 \\        
FCC147 	      & 1.376 $\pm$ 0.014 & 29.543 $\pm$ 0.023 & 31.458 $\pm$ 0.070 & 19.6 $\pm$ 0.6  &  11.9  & NGC1374 \\        
FCC148 	      & 1.225 $\pm$ 0.009 & 29.367 $\pm$ 0.037 & 31.500 $\pm$ 0.072 & 19.9 $\pm$ 0.7  &  13.6  & NGC1375 \\        
FCC152 	      & 1.125 $\pm$ 0.011 & 29.130 $\pm$ 0.021 & 31.357 $\pm$ 0.065 & 18.7 $\pm$ 0.6  &  14.1  & ESO358-25 \\      
FCC153 	      & 1.262 $\pm$ 0.009 & 29.498 $\pm$ 0.034 & 31.588 $\pm$ 0.071 & 20.8 $\pm$ 0.7  &  13.0  & ESO358-26 \\      
FCC167 	      & 1.394 $\pm$ 0.019 & 29.750 $\pm$ 0.021 & 31.632 $\pm$ 0.075 & 21.2 $\pm$ 0.7  &  11.3  & NGC1380 \\        
FCC170 	      & 1.376 $\pm$ 0.019 & 29.790 $\pm$ 0.028 & 31.705 $\pm$ 0.076 & 21.9 $\pm$ 0.8  &  13.0  & NGC1381 \\        
FCC177 	      & 1.257 $\pm$ 0.009 & 29.412 $\pm$ 0.019 & 31.509 $\pm$ 0.065 & 20.0 $\pm$ 0.6  &  13.2  & NGC1380A \\       
FCC182 	      & 1.302 $\pm$ 0.029 & 29.421 $\pm$ 0.044 & 31.458 $\pm$ 0.086 & 19.6 $\pm$ 0.8  &  14.9  & \dots \\          
FCC184 	      & 1.517 $\pm$ 0.011 & 29.857 $\pm$ 0.052 & 31.430 $\pm$ 0.087 & 19.3 $\pm$ 0.8  &  12.3  & NGC1387 \\        
FCC190 	      & 1.352 $\pm$ 0.016 & 29.581 $\pm$ 0.029 & 31.540 $\pm$ 0.073 & 20.3 $\pm$ 0.7  &  13.5  & NGC1382\tmk{a} \\ 
FCC193 	      & 1.369 $\pm$ 0.010 & 29.699 $\pm$ 0.033 & 31.627 $\pm$ 0.072 & 21.2 $\pm$ 0.7  &  12.8  & NGC1389 \\        
FCC202 	      & 1.157 $\pm$ 0.049 & 29.312 $\pm$ 0.036 & 31.511 $\pm$ 0.083 & 20.1 $\pm$ 0.8  &  15.3  & NGC1396 \\        
FCC203 	      & 1.169 $\pm$ 0.010 & 29.390 $\pm$ 0.055 & 31.579 $\pm$ 0.082 & 20.7 $\pm$ 0.8  &  15.5  & ESO358-42 \\      
FCC204 	      & 1.233 $\pm$ 0.011 & 29.495 $\pm$ 0.043 & 31.619 $\pm$ 0.075 & 21.1 $\pm$ 0.7  &  14.9  & ESO358-43 \\      
FCC213 	      & 1.500 $\pm$ 0.022 & 29.972 $\pm$ 0.020 & 31.596 $\pm$ 0.091 & 20.9 $\pm$ 0.9  &  10.6  & NGC1399 \\        
FCC219 	      & 1.479 $\pm$ 0.013 & 29.842 $\pm$ 0.016 & 31.526 $\pm$ 0.072 & 20.2 $\pm$ 0.7  &  10.9  & NGC1404 \\        
FCC249 	      & 1.295 $\pm$ 0.022 & 29.752 $\pm$ 0.045 & 31.799 $\pm$ 0.082 & 22.9 $\pm$ 0.9  &  13.6  & NGC1419 \\        
FCC255 	      & 1.197 $\pm$ 0.013 & 29.340 $\pm$ 0.023 & 31.502 $\pm$ 0.067 & 20.0 $\pm$ 0.6  &  13.7  & ESO358-50 \\      
FCC276 	      & 1.366 $\pm$ 0.007 & 29.524 $\pm$ 0.026 & 31.459 $\pm$ 0.068 & 19.6 $\pm$ 0.6  &  11.8  & NGC1427 \\        
FCC277 	      & 1.313 $\pm$ 0.011 & 29.558 $\pm$ 0.046 & 31.579 $\pm$ 0.078 & 20.7 $\pm$ 0.7  &  13.8  & NGC1428 \\        
FCC288 	      & 1.124 $\pm$ 0.009 & 29.453 $\pm$ 0.050 & 31.680 $\pm$ 0.079 & 21.7 $\pm$ 0.8  &  15.4  & ESO358-56 \\      
FCC301 	      & 1.232 $\pm$ 0.013 & 29.348 $\pm$ 0.047 & 31.473 $\pm$ 0.078 & 19.7 $\pm$ 0.7  &  14.2  & ESO358-59 \\      
FCC303	      & 1.143 $\pm$ 0.017 & 29.261 $\pm$ 0.043 & 31.471 $\pm$ 0.076 & 19.7 $\pm$ 0.7  &  15.5  & \dots \\ 
FCC310 	      & 1.314 $\pm$ 0.007 & 29.479 $\pm$ 0.020 & 31.499 $\pm$ 0.065 & 19.9 $\pm$ 0.6  &  13.5  & NGC1460 \\        
FCC324 	      & 1.156 $\pm$ 0.010 & 29.314 $\pm$ 0.044 & 31.514 $\pm$ 0.076 & 20.1 $\pm$ 0.7  &  15.3  & ESO358-66 \\      
FCC335 	      & 1.131 $\pm$ 0.013 & 29.220 $\pm$ 0.029 & 31.442 $\pm$ 0.068 & 19.4 $\pm$ 0.6  &  14.2  & ESO359-02 \\      
NGC1340       & 1.314 $\pm$ 0.007 & 29.583 $\pm$ 0.028 & 31.603 $\pm$ 0.068 & 20.9 $\pm$ 0.7  &  11.3  & NGC1344\tmk{b} \\ 
IC2006        & 1.422 $\pm$ 0.016 & 29.702 $\pm$ 0.049 & 31.525 $\pm$ 0.086 & 20.2 $\pm$ 0.8  &  12.2  & ESO359-07 \\      
\enddata
\tablenotetext{a}{Also known as NGC1380B.}
\tablenotetext{b}{Due to an apparent error in the catalogue, this galaxy has two NGC numbers.}
\tablenotetext{c}{The distance calibration is less well constrained at these blue colors.}
\tablecomments{Columns list:  (1)~FCC designation (Ferguson 1989); 
  (2)~mean \gz\ color of the analyzed region;
  (3)~mean SBF magnitude \mbarz;
  (4)~distance modulus derived from equation\,\ref{eq:cal}; 
  (5)~distance in Mpc; 
  (6)~total $B$~magnitude from the FCC or NED; 
  (7)~common galaxy name. 
The $B_T$ column facilitates comparison to tables in other papers of this series where
the galaxies are ordered by this quantity.}
\end{deluxetable}

\newpage

\def\tmk#1{\tablenotemark{#1}}
% table generated by textab.py !
\begin{deluxetable}{lrrrrrc}
%\tabletypesize{\small}
\tabletypesize{\footnotesize}
\tablecaption{SBF Data for Virgo Galaxies\label{tab:virgo}}
\tablewidth{0pt}
\tablehead{
\colhead{Galaxy} &
\colhead{$(g{-}z)$} &
\colhead{$\mbarz$} &
\colhead{\mM} &
\colhead{$d$} &
\colhead{$B_T$} &
\colhead{Name} \\
\colhead{(1)} &
\colhead{(2)} &
\colhead{(3)} &
\colhead{(4)} &
\colhead{(5)} &
\colhead{(6)} &
\colhead{(7)}
}
\startdata
  VCC9         & 1.054 $\pm$ 0.014 & 28.885 $\pm$ 0.066 & 31.170 $\pm$ 0.091 & 17.1 $\pm$ 0.7  &  13.9  & IC3019 \\        
 VCC21\tmk{c}  & 0.898 $\pm$ 0.015 & 28.993 $\pm$ 0.040 & 31.421 $\pm$ 0.130 & 19.2 $\pm$ 1.1  &  14.8  & IC3025 \\        
 VCC33 	       & 1.014 $\pm$ 0.014 & 28.568 $\pm$ 0.051 & 30.886 $\pm$ 0.133 & 15.0 $\pm$ 0.9  &  14.7  & IC3032 \\        
VCC140 	       & 1.125 $\pm$ 0.014 & 28.857 $\pm$ 0.040 & 31.084 $\pm$ 0.074 & 16.5 $\pm$ 0.6  &  14.3  & IC3065 \\        
VCC200 	       & 1.164 $\pm$ 0.014 & 29.114 $\pm$ 0.050 & 31.307 $\pm$ 0.080 & 18.3 $\pm$ 0.7  &  14.7  & \dots \\         
VCC230 	       & 1.156 $\pm$ 0.014 & 29.063 $\pm$ 0.091 & 31.263 $\pm$ 0.110 & 17.9 $\pm$ 0.9  &  15.2  & IC3101 \\        
VCC355 	       & 1.404 $\pm$ 0.014 & 29.089 $\pm$ 0.024 & 30.950 $\pm$ 0.072 & 15.5 $\pm$ 0.5  &  12.4  & NGC4262 \\       
VCC369 	       & 1.440 $\pm$ 0.014 & 29.215 $\pm$ 0.022 & 30.996 $\pm$ 0.073 & 15.8 $\pm$ 0.5  &  11.8  & NGC4267 \\       
VCC437 	       & 1.208 $\pm$ 0.014 & 29.029 $\pm$ 0.049 & 31.180 $\pm$ 0.080 & 17.2 $\pm$ 0.6  &  14.5  & UGC7399A \\      
VCC538\tmk{d}  & 1.110 $\pm$ 0.014 & 29.572 $\pm$ 0.064 & 31.812 $\pm$ 0.089 & 23.0 $\pm$ 0.9  &  15.4  & NGC4309A \\      
VCC543 	       & 1.163 $\pm$ 0.014 & 28.795 $\pm$ 0.045 & 30.989 $\pm$ 0.077 & 15.8 $\pm$ 0.6  &  14.4  & UGC7436 \\       
VCC571\tmk{d}  & 1.063 $\pm$ 0.014 & 29.607 $\pm$ 0.078 & 31.885 $\pm$ 0.100 & 23.8 $\pm$ 1.1  &  14.7  & \dots \\         
VCC575\tmk{d}  & 1.272 $\pm$ 0.014 & 29.639 $\pm$ 0.041 & 31.716 $\pm$ 0.076 & 22.0 $\pm$ 0.8  &  14.1  & NGC4318 \\       
VCC698 	       & 1.298 $\pm$ 0.014 & 29.294 $\pm$ 0.032 & 31.337 $\pm$ 0.072 & 18.5 $\pm$ 0.6  &  13.6  & NGC4352 \\       
VCC731\tmk{d}  & 1.489 $\pm$ 0.014 & 30.161 $\pm$ 0.015 & 31.816 $\pm$ 0.074 & 23.1 $\pm$ 0.8  &  10.5  & NGC4365 \\       
VCC751 	       & 1.253 $\pm$ 0.014 & 28.891 $\pm$ 0.046 & 30.992 $\pm$ 0.078 & 15.8 $\pm$ 0.6  &  15.3  & IC3292 \\        
VCC759 	       & 1.460 $\pm$ 0.014 & 29.407 $\pm$ 0.018 & 31.139 $\pm$ 0.073 & 16.9 $\pm$ 0.6  &  11.8  & NGC4371 \\       
VCC763 	       & 1.431 $\pm$ 0.014 & 29.535 $\pm$ 0.013 & 31.337 $\pm$ 0.070 & 18.5 $\pm$ 0.6  &  10.3  & NGC4374,\,M84 \\ 
VCC778 	       & 1.320 $\pm$ 0.014 & 29.225 $\pm$ 0.030 & 31.236 $\pm$ 0.071 & 17.7 $\pm$ 0.6  &  12.7  & NGC4377 \\       
VCC784 	       & 1.368 $\pm$ 0.014 & 29.073 $\pm$ 0.024 & 31.004 $\pm$ 0.070 & 15.9 $\pm$ 0.5  &  12.7  & NGC4379 \\       
VCC798 	       & 1.349 $\pm$ 0.014 & 29.298 $\pm$ 0.013 & 31.262 $\pm$ 0.067 & 17.9 $\pm$ 0.5  &  10.1  & NGC4382,\,M85 \\ 
VCC828 	       & 1.369 $\pm$ 0.014 & 29.347 $\pm$ 0.031 & 31.276 $\pm$ 0.073 & 18.0 $\pm$ 0.6  &  12.8  & NGC4387 \\       
VCC856 	       & 1.164 $\pm$ 0.014 & 28.947 $\pm$ 0.044 & 31.140 $\pm$ 0.076 & 16.9 $\pm$ 0.6  &  14.2  & IC3328 \\        
VCC881\tmk{a}  & 1.460 $\pm$ 0.014 & 29.529 $\pm$ 0.013 & 31.261 $\pm$ 0.072 & 17.9 $\pm$ 0.6  &  10.1  & NGC4406,\,M86 \\ 
VCC944         & 1.378 $\pm$ 0.014 & 29.112 $\pm$ 0.021 & 31.024 $\pm$ 0.070 & 16.0 $\pm$ 0.5  &  12.1  & NGC4417 \\       
VCC1025\tmk{d} & 1.380 $\pm$ 0.015 & 29.850 $\pm$ 0.030 & 31.759 $\pm$ 0.074 & 22.5 $\pm$ 0.8  &  13.1  & NGC4434 \\       
VCC1030        & 1.306 $\pm$ 0.014 & 29.076 $\pm$ 0.020 & 31.107 $\pm$ 0.067 & 16.7 $\pm$ 0.5  &  11.8  & NGC4435 \\       
VCC1049        & 1.051 $\pm$ 0.014 & 28.736 $\pm$ 0.045 & 31.023 $\pm$ 0.077 & 16.0 $\pm$ 0.6  &  14.2  & UGC7580 \\       
VCC1062        & 1.441 $\pm$ 0.014 & 29.146 $\pm$ 0.018 & 30.925 $\pm$ 0.072 & 15.3 $\pm$ 0.5  &  11.4  & NGC4442 \\       
VCC1075        & 1.150 $\pm$ 0.014 & 28.848 $\pm$ 0.076 & 31.054 $\pm$ 0.098 & 16.2 $\pm$ 0.7  &  15.1  & IC3383 \\        
VCC1087        & 1.236 $\pm$ 0.014 & 28.989 $\pm$ 0.040 & 31.110 $\pm$ 0.075 & 16.7 $\pm$ 0.6  &  14.3  & IC3381 \\        
VCC1146        & 1.274 $\pm$ 0.014 & 28.988 $\pm$ 0.028 & 31.063 $\pm$ 0.070 & 16.3 $\pm$ 0.5  &  12.9  & NGC4458 \\       
VCC1154        & 1.462 $\pm$ 0.014 & 29.295 $\pm$ 0.020 & 31.022 $\pm$ 0.073 & 16.0 $\pm$ 0.5  &  11.4  & NGC4459 \\       
VCC1178        & 1.370 $\pm$ 0.014 & 29.066 $\pm$ 0.031 & 30.993 $\pm$ 0.073 & 15.8 $\pm$ 0.5  &  13.4  & NGC4464 \\       
VCC1185        & 1.239 $\pm$ 0.014 & 29.028 $\pm$ 0.089 & 31.145 $\pm$ 0.109 & 16.9 $\pm$ 0.9  &  15.7  & \dots \\         
VCC1226        & 1.518 $\pm$ 0.014 & 29.546 $\pm$ 0.011 & 31.116 $\pm$ 0.075 & 16.7 $\pm$ 0.6  &   9.3  & NGC4472,\,M49 \\ 
VCC1231        & 1.446 $\pm$ 0.014 & 29.149 $\pm$ 0.017 & 30.916 $\pm$ 0.072 & 15.2 $\pm$ 0.5  &  11.1  & NGC4473 \\       
VCC1242        & 1.306 $\pm$ 0.014 & 28.915 $\pm$ 0.048 & 30.946 $\pm$ 0.080 & 15.5 $\pm$ 0.6  &  12.6  & NGC4474 \\       
VCC1250        & 1.145 $\pm$ 0.014 & 29.027 $\pm$ 0.049 & 31.237 $\pm$ 0.079 & 17.7 $\pm$ 0.6  &  12.9  & NGC4476 \\       
VCC1261        & 1.131 $\pm$ 0.014 & 29.076 $\pm$ 0.035 & 31.298 $\pm$ 0.071 & 18.2 $\pm$ 0.6  &  13.6  & NGC4482 \\       
VCC1279        & 1.402 $\pm$ 0.014 & 29.294 $\pm$ 0.023 & 31.159 $\pm$ 0.071 & 17.1 $\pm$ 0.6  &  12.2  & NGC4478 \\       
VCC1283        & 1.385 $\pm$ 0.014 & 29.304 $\pm$ 0.030 & 31.203 $\pm$ 0.073 & 17.4 $\pm$ 0.6  &  13.4  & NGC4479 \\       
VCC1297        & 1.457 $\pm$ 0.014 & 29.315 $\pm$ 0.045 & 31.055 $\pm$ 0.084 & 16.3 $\pm$ 0.6  &  14.3  & NGC4486B \\      
VCC1303        & 1.354 $\pm$ 0.014 & 29.164 $\pm$ 0.027 & 31.120 $\pm$ 0.071 & 16.7 $\pm$ 0.5  &  13.1  & NGC4483 \\       
VCC1316        & 1.526 $\pm$ 0.014 & 29.566 $\pm$ 0.012 & 31.111 $\pm$ 0.076 & 16.7 $\pm$ 0.6  &   9.6  & NGC4486,\,M87 \\ 
VCC1321        & 1.259 $\pm$ 0.014 & 28.841 $\pm$ 0.028 & 30.935 $\pm$ 0.069 & 15.4 $\pm$ 0.5  &  12.8  & NGC4489 \\       
VCC1327        & 1.400 $\pm$ 0.014 & 29.450 $\pm$ 0.030 & 31.319 $\pm$ 0.074 & 18.4 $\pm$ 0.6  &  13.3  & NGC4486A \\      
VCC1355        & 1.115 $\pm$ 0.014 & 28.910 $\pm$ 0.063 & 31.145 $\pm$ 0.088 & 16.9 $\pm$ 0.7  &  14.3  & IC3442 \\        
VCC1407        & 1.220 $\pm$ 0.014 & 28.990 $\pm$ 0.030 & 31.128 $\pm$ 0.070 & 16.8 $\pm$ 0.5  &  15.5  & IC3461 \\        
VCC1422        & 1.180 $\pm$ 0.014 & 28.759 $\pm$ 0.039 & 30.937 $\pm$ 0.074 & 15.4 $\pm$ 0.5  &  13.6  & IC3468 \\        
VCC1431        & 1.280 $\pm$ 0.014 & 28.958 $\pm$ 0.042 & 31.025 $\pm$ 0.076 & 16.0 $\pm$ 0.6  &  14.5  & IC3470 \\        
VCC1440        & 1.186 $\pm$ 0.014 & 28.858 $\pm$ 0.048 & 31.030 $\pm$ 0.079 & 16.1 $\pm$ 0.6  &  15.2  & IC798 \\         
VCC1475        & 1.215 $\pm$ 0.014 & 28.968 $\pm$ 0.032 & 31.111 $\pm$ 0.070 & 16.7 $\pm$ 0.5  &  13.4  & NGC4515 \\       
VCC1488\tmk{c} & 0.850 $\pm$ 0.014 & 28.566 $\pm$ 0.053 & 31.053 $\pm$ 0.134 & 16.2 $\pm$ 1.0  &  14.8  & IC3487 \\        
VCC1489\tmk{c} & 0.993 $\pm$ 0.014 & 28.964 $\pm$ 0.097 & 31.299 $\pm$ 0.156 & 18.2 $\pm$ 1.3  &  15.9  & IC3490 \\        
VCC1512        & 1.276 $\pm$ 0.014 & 29.240 $\pm$ 0.030 & 31.312 $\pm$ 0.070 & 18.3 $\pm$ 0.6  &  15.7  & \dots \\         
VCC1528        & 1.218 $\pm$ 0.014 & 28.926 $\pm$ 0.043 & 31.066 $\pm$ 0.076 & 16.3 $\pm$ 0.6  &  14.5  & IC3501 \\        
VCC1537        & 1.300 $\pm$ 0.014 & 28.936 $\pm$ 0.029 & 30.976 $\pm$ 0.070 & 15.7 $\pm$ 0.5  &  12.7  & NGC4528 \\       
VCC1539        & 1.147 $\pm$ 0.014 & 28.939 $\pm$ 0.102 & 31.147 $\pm$ 0.119 & 17.0 $\pm$ 0.9  &  15.7  & \dots \\         
VCC1545        & 1.253 $\pm$ 0.014 & 29.023 $\pm$ 0.050 & 31.124 $\pm$ 0.081 & 16.8 $\pm$ 0.6  &  15.0  & IC3509 \\        
VCC1619        & 1.298 $\pm$ 0.014 & 28.886 $\pm$ 0.022 & 30.929 $\pm$ 0.068 & 15.3 $\pm$ 0.5  &  12.5  & NGC4550 \\       
VCC1627\tmk{b} & 1.326 $\pm$ 0.130 & 28.962 $\pm$ 0.049 & 30.963 $\pm$ 0.217 & 15.6 $\pm$ 1.6  &  15.2  & \dots \\         
VCC1630        & 1.417 $\pm$ 0.014 & 29.212 $\pm$ 0.028 & 31.045 $\pm$ 0.074 & 16.2 $\pm$ 0.5  &  12.9  & NGC4551 \\       
VCC1632\tmk{a} & 1.488 $\pm$ 0.014 & 29.363 $\pm$ 0.013 & 31.021 $\pm$ 0.073 & 16.0 $\pm$ 0.5  &  10.8  & NGC4552,\,M89 \\ 
VCC1661        & 1.255 $\pm$ 0.014 & 28.901 $\pm$ 0.156 & 31.000 $\pm$ 0.168 & 15.8 $\pm$ 1.2  &  16.0  & \dots \\         
VCC1664        & 1.421 $\pm$ 0.014 & 29.188 $\pm$ 0.021 & 31.013 $\pm$ 0.071 & 15.9 $\pm$ 0.5  &  12.0  & NGC4564 \\       
VCC1692        & 1.421 $\pm$ 0.014 & 29.343 $\pm$ 0.020 & 31.168 $\pm$ 0.071 & 17.1 $\pm$ 0.6  &  11.8  & NGC4570 \\       
VCC1695        & 1.055 $\pm$ 0.014 & 28.809 $\pm$ 0.050 & 31.093 $\pm$ 0.080 & 16.5 $\pm$ 0.6  &  14.5  & IC3586 \\        
VCC1720        & 1.410 $\pm$ 0.014 & 29.220 $\pm$ 0.020 & 31.068 $\pm$ 0.071 & 16.4 $\pm$ 0.5  &  12.3  & NGC4578 \\       
VCC1743        & 1.073 $\pm$ 0.014 & 28.964 $\pm$ 0.071 & 31.234 $\pm$ 0.094 & 17.6 $\pm$ 0.8  &  15.5  & IC3602 \\        
VCC1779\tmk{c} & 0.904 $\pm$ 0.014 & 28.420 $\pm$ 0.055 & 30.841 $\pm$ 0.135 & 14.7 $\pm$ 0.9  &  14.8  & IC3612 \\        
VCC1826        & 1.117 $\pm$ 0.014 & 28.826 $\pm$ 0.060 & 31.060 $\pm$ 0.086 & 16.3 $\pm$ 0.6  &  15.7  & IC3633 \\        
VCC1828        & 1.183 $\pm$ 0.014 & 28.970 $\pm$ 0.059 & 31.145 $\pm$ 0.086 & 16.9 $\pm$ 0.7  &  15.3  & IC3635 \\        
VCC1833        & 1.139 $\pm$ 0.014 & 28.844 $\pm$ 0.041 & 31.059 $\pm$ 0.074 & 16.3 $\pm$ 0.6  &  14.5  & \dots \\         
VCC1861        & 1.260 $\pm$ 0.014 & 28.941 $\pm$ 0.045 & 31.034 $\pm$ 0.078 & 16.1 $\pm$ 0.6  &  14.4  & IC3652 \\        
VCC1871        & 1.356 $\pm$ 0.014 & 28.998 $\pm$ 0.039 & 30.950 $\pm$ 0.076 & 15.5 $\pm$ 0.5  &  13.9  & IC3653 \\        
VCC1883        & 1.274 $\pm$ 0.014 & 29.017 $\pm$ 0.023 & 31.092 $\pm$ 0.068 & 16.5 $\pm$ 0.5  &  12.6  & NGC4612 \\       
VCC1886\tmk{c} & 0.951 $\pm$ 0.014 & 28.612 $\pm$ 0.137 & 30.985 $\pm$ 0.184 & 15.7 $\pm$ 1.3  &  15.5  & \dots \\         
VCC1895        & 1.116 $\pm$ 0.014 & 28.770 $\pm$ 0.030 & 31.005 $\pm$ 0.069 & 15.9 $\pm$ 0.5  &  14.9  & UGC7854 \\       
VCC1903        & 1.458 $\pm$ 0.014 & 29.122 $\pm$ 0.014 & 30.859 $\pm$ 0.072 & 14.9 $\pm$ 0.5  &  10.8  & NGC4621,\,M59 \\ 
VCC1910        & 1.331 $\pm$ 0.014 & 29.033 $\pm$ 0.034 & 31.027 $\pm$ 0.073 & 16.0 $\pm$ 0.5  &  14.2  & IC809 \\         
VCC1913        & 1.332 $\pm$ 0.014 & 29.199 $\pm$ 0.028 & 31.191 $\pm$ 0.071 & 17.3 $\pm$ 0.6  &  13.2  & NGC4623 \\       
VCC1938        & 1.305 $\pm$ 0.014 & 29.156 $\pm$ 0.024 & 31.189 $\pm$ 0.069 & 17.3 $\pm$ 0.5  &  12.1  & NGC4638 \\       
VCC1948\tmk{c} & 0.992 $\pm$ 0.014 & 28.804 $\pm$ 0.099 & 31.140 $\pm$ 0.158 & 16.9 $\pm$ 1.2  &  15.1  & \dots \\         
VCC1978        & 1.559 $\pm$ 0.014 & 29.647 $\pm$ 0.012 & 31.082 $\pm$ 0.079 & 16.5 $\pm$ 0.6  &   9.8  & NGC4649,\,M60 \\ 
VCC1993        & 1.230 $\pm$ 0.014 & 28.970 $\pm$ 0.030 & 31.097 $\pm$ 0.070 & 16.6 $\pm$ 0.5  &  15.3  & \dots \\         
VCC2000        & 1.336 $\pm$ 0.014 & 28.890 $\pm$ 0.024 & 30.876 $\pm$ 0.069 & 15.0 $\pm$ 0.5  &  11.9  & NGC4660 \\       
VCC2019        & 1.154 $\pm$ 0.014 & 28.970 $\pm$ 0.045 & 31.172 $\pm$ 0.077 & 17.2 $\pm$ 0.6  &  14.6  & IC3735 \\        
VCC2050        & 1.112 $\pm$ 0.014 & 28.762 $\pm$ 0.056 & 31.000 $\pm$ 0.084 & 15.8 $\pm$ 0.6  &  15.2  & IC3779 \\        
VCC2092        & 1.464 $\pm$ 0.014 & 29.306 $\pm$ 0.018 & 31.028 $\pm$ 0.073 & 16.1 $\pm$ 0.5  &  11.5  & NGC4754 \\       
NGC4697\tmk{e} & 1.375 $\pm$ 0.008 & 28.573 $\pm$ 0.019 & 30.491 $\pm$ 0.065 & 12.5 $\pm$ 0.4  &  10.1  & NGC4697 \\ 
\enddata
\tablenotetext{a}{There were typographical errors in the 
  data for these galaxies in Table~1 of ACSVCS-XIII.}
\tablenotetext{b}{This galaxy had incorrect distance moduli listed in
  Table~5 of ACSVCS-XIII.}
\tablenotetext{c}{The distance calibration is less well constrained at these blue colors.}
\tablenotetext{d}{Virgo \wprime\ group galaxy.}
\tablenotetext{e}{Member of a group in the Virgo Southern Extension; data described in \jordan\ \etal\ (2005).}
\tablecomments{Columns list:  (1) VCC designation (Binggeli \etal\ 1985); 
  (2)~mean \gz\ color of the analyzed region from \vcsxiii\ (Mei et al.\ 2007);
  (3)~mean SBF magnitude \mbarz\ from \vcsxiii;
  (4)~distance modulus derived from equation\,\ref{eq:cal}; 
  (5)~distance in Mpc; 
  (6)~total $B$~magnitude from the VCC or NED; 
  (7)~common galaxy name. 
The $B_T$ column facilitates comparison to tables in papers of the ACSVCS series where
the galaxies are ordered by this quantity.}
\end{deluxetable}

\end{document}